\documentclass[10pt,a4paper]{article} 
\pdfoutput=1
\usepackage{a4wide}
\usepackage{graphicx} 
\usepackage{amsmath} 
\usepackage{amssymb}
\newtheorem{Def}{Definition} 
\newtheorem{The}{Theorem}
\newtheorem{Pro}{Proof}
\usepackage[round,longnamesfirst]{natbib}
\begin{document} 
\title{A Model-Based Frequency Constraint for Mining Associations
from Transaction Data}
\author{Michael~Hahsler 
\hfill \small  michael.hahsler@wu-wien.ac.at \\ 
\small Vienna
University of Economics and Business Administration
}
\date{12 May 2006}
\maketitle
\begin{abstract} 
Mining frequent itemsets is a popular method for finding
associated items in databases.
For this method, support, the co-occurrence frequency of the items
which form an association, is used as the primary indicator of the
associations's significance. A single user-specified support threshold is
used to decided if associations should be further investigated.
Support has some known problems with rare items, favors shorter itemsets
and sometimes produces misleading associations.

In this paper we develop a novel model-based frequency constraint 
as an alternative to a single, user-specified minimum support.
The constraint utilizes knowledge of the process generating transaction data
by applying a simple stochastic mixture model (the NB model) which
allows for transaction data's typically highly skewed item 
frequency distribution.
A user-specified precision threshold is used together with the model to
find local frequency thresholds for groups of itemsets.
Based on the constraint we develop the notion of NB-frequent
itemsets and adapt a mining algorithm to find all
NB-frequent itemsets in a database.  In experiments with publicly
available transaction databases we show that the new constraint
provides improvements over a single  minimum support
threshold and that the precision threshold is more robust and 
easier to set and interpret by the user.  
\end{abstract}
\section{Introduction} 
Mining associations (i.e., set of associated items) in large
databases has been under intense research since 
\cite{Agrawal1993} presented
{\em Apriori}, the first algorithm using the support-con\-fidence
framework to mine frequent itemsets and association rules.
The enormous interest in associations between items
is due
to their direct applicability for many practical purposes.
Beginning with discovering regularities in transaction data
recorded by point-of-sale systems to improve sales, associations 
are also used to analyze Web usage patterns~\citep{Srivastava2000},
for intrusion detection~\citep{Luo2000}, 
for mining genome data~\citep{Creighton2003}, 
and for many other applications.

An association is a set of items
found in a database which provides
useful and actionable insights into the structure of the data.
For most current applications support is used to find 
potentially useful associations.
Support is a measure of significance defined as the relative 
frequency of an association in the database.
The main advantages of using support are that it is simple to calculate,
no assumptions about the structure of mined data are required, and
support possesses the a so-called 
{\em downward closure property}~\citep{Agrawal1994} 
which makes a more efficient search for all frequent itemsets 
in a database possible. 
However, support has also some important shortcomings.
Some examples found in the literature are:
\begin{itemize}
\item
\cite{Silverstein1998} argue with the help of examples
that the definition of association
rules (using support and confidence) can produce misleading associations.
The authors suggest using statistical tests instead of support
to find reliable dependencies between items.
\item 
Support is prone to the rare item problem~\citep{Liu1999}
where associations including items with low support are discarded
although they might contain valuable information.
\item
Support favors smaller itemsets while longer itemsets
could still be interesting, even if they are less frequent~\citep{Seno2001}. 
In order to find longer itemset, one would have to lower the support threshold
which would lead to an explosion of the number of short itemsets found.
\end{itemize}

Statistics provides a multitude of
models which proved to be extremely helpful to describe data
frequently mined for associations (e.g., accident data,
market research data including market baskets, data from medical
and military applications, and biometrical data~\citep{Johnson1993}).
For transaction data, many models build on mixtures of counting processes 
which are known to result in extremely skewed item frequency distributions
with very few relatively frequent items while most items are infrequent.
This is especially problematic since support's
rare item problem affects the majority of items
in such a database.
Although the effects of skewed item frequency distributions in
transaction data are sometimes discussed (e.g. by~\cite{Liu1999} 
or \cite{Xiong2003}),
most current approaches neglect knowledge
about statistical properties of the generating processes 
which underlie the mined databases.

The contribution of this paper is that we
address the shortcomings of a single, user-specified minimum support
threshold by departing from finding frequent itemsets.
Instead we propose a model-based frequency constraint to find
NB-frequent itemsets.
For this constraint we utilizes knowledge of the process which underlies
transaction data by applying a simple stochastic baseline model 
(an extension of the NB model) which is
known for its wide applicability.  
A user-specified precision threshold is used to identify 
local frequency thresholds for groups of associations based on
evaluating observed deviations from a baseline model.
The proposed model-based constraint has the following properties: 

\begin{enumerate}
\item It reduces the problem
with rare items since the used stochastic model allows for highly
skewed frequency distributions.  
\item It is able to produce longer
associations without generating an enormous number of shorter, spurious
associations since the support required by the model 
is set locally and decreases with the number of items
forming an association.  
\item Its precision threshold parameter can be interpreted as a predicted error
rate.  This makes communicating and setting the parameter easier for domain
experts.  Also, the parameter seems to be less dependent on the structure of
the database than support.
\end{enumerate}

The rest of the paper is organized as follows: 
In the next section we review the background of mining
associations and some proposed alternative frequency constraints.  
In Section~\ref{development} we develop the model-based frequency
constraint, the concept of NB-frequent itemsets, and show that the
chosen model is useful to describe real-word transaction data.  
In Section~\ref{algorithms} we present an algorithm to mine
all NB-frequent itemsets in a database.
In Section~\ref{evaluation} we investigate and discuss the behavior
of the model-based constraint using several real-world and
artificial transaction data sets.

\section{Background and Related Work}
The problem of mining associated items (frequent itemsets) from transaction
data was formally introduced by \cite{Agrawal1993}
for mining association rules as: 
Let $I=\{i_1, i_2,...,i_n\}$ be a set of $n$ distinct literals
called items and ${\cal D}=\{t_1, t_2,...,t_m\}$ a set of
transactions called the database. Each transaction in ${\cal D}$
contains a subset of the items in $I$. A rule is defined as an
implication of the from $X \longrightarrow Y$ where $X,Y \subseteq I$
and $X \cap Y = \emptyset$.  The sets of items (for short itemsets)
$X$ and $Y$  are called antecedent and consequent of the rule.
An itemset which contains $k$ items is said to have length or size $k$ 
and is called a $k$-itemset.
An itemset which is produced by adding a single item to another
itemset is called a $1$-extension of the latter itemset.

Constraints on various measures of significance and interest can be
used to select interesting associations and rules. 
\cite{Agrawal1993} define the measures
support and confidence for association rules.

\begin{Def}[Support] Support is defined on itemset $Z \subseteq I$ as the
proportion of transactions in which all items in $Z$ are 
found together in the database:

\begin{displaymath} 
\mathrm{supp}(Z) = 
\frac{\mathrm{freq}(Z)}{|{\cal D}|}, 
\label{supp} 
\end{displaymath}

where $\mathrm{freq}(Z)$ denotes the frequency of itemset $Z$ 
(number of transactions in which $Z$ occurs)
in database ${\cal D}$, and $|{\cal D}|$ 
is the number of transactions in the database.
\end{Def}

 %
%
%
Confidence is defined for a rule $X \longrightarrow Y$ as the ratio
$\mathrm{supp}(X \cup Y)/\mathrm{supp}(X)$. Since confidence is not
a frequency constraints we will only discuss support in
the following.

An itemset $Z$ is only considered 
significant and interesting in the association rule framework if
the constraint $\mathrm{supp}(Z) \ge \sigma$ holds, where $\sigma$
is a user-specified minimum support.  Itemsets which satisfy the
minimum support constraint are called {\em frequent itemsets} since
their occurrence frequency surpasses a set frequency threshold,
hence the name frequency constraint.  Some authors refer to
frequent itemsets also as {\em large itemsets}~\citep{Agrawal1993} or
{\em covering sets}~\citep{Mannila1994}.

The rational for minimum support is that items which appear more
often in the database are more important since, e.g. in a sales
setting they are responsible for a higher sales volume.  However,
this rational breaks down when some rare but expensive items
contribute most to the store's overall earnings.  Not finding
associations for such items is known as support's rare item
problem~\citep{Liu1999}.  
Support also systematically favors smaller itemsets~\citep{Seno2001}.  
By adding items to
an itemset the probability of finding such longer itemsets in the
database can only decrease or, in rare cases, stay the same.
Consequently, longer itemsets are less likely to meet the 
minimum support. Reducing minimum support to find longer itemsets normally
results in an explosion of the number of small itemsets found, which
makes this approach infeasible for most applications.

For all but very small or extremely sparse databases, finding all
frequent itemsets is computationally very expensive since
the search space for frequent itemsets
grows exponentially with the number of items. 
However, the minimum support constraint possesses a special
property called {\em downward closure}~\citep{Agrawal1994} 
(also called {\em anti-monotonicity}~\citep{Pei2001}) 
which can be used to make more efficient search possible.
%
A constraint is downward closed
(anti-monotone) if, and only if, for each itemset which satisfies
the constraint all subsets also satisfy the constraint.  
%
The frequency constraint  minimum support is downward closed since
if set $X$ is supported at a threshold $\sigma$, also all its
subsets $Y \subset X$, which can only have a higher or the same
support as $X$, must be supported at the same threshold.  This
property implies that 
(a) an itemset can only satisfy a downward
closed constraint if all its subsets satisfy the constraint and
that 
(b) if an itemset is found to satisfy a downward closed
constraint all its subsets need no inspection since they must also
satisfy the constraint.  
These facts are used by mining algorithms
to reduce the search space which is often referred to as 
{\em pruning} or finding a border in the 
lattice representation of the
search space.

Driven by support's problems with rare items and skewed item
frequency distributions, some researchers proposed alternatives for
mining associations. In the following we will 
review some approaches which are related to this work.

\cite{Liu1999} try to alleviate the rare item problem. They
suggest mining itemsets with individual 
{\em minimum item support} thresholds assigned to each item.
\citeauthor{Liu1999} showed that after sorting the items according
to their minimum item support a {\em sorted closure property} of
minimum item support can be used to prune the search space.  
A open research question is how to determine the optimal values
for the minimum item supports, especially in databases with many
items where a manual assignment is not feasible.

\cite{Seno2001} try to reduce support's tendency
to favor smaller itemsets by proposing a minimum support which
decreases as a function of itemset length.
Since this invalidates the downward closure of support,
the authors develop a property called 
{\em smallest valid extension,} which
can be exploited for pruning the search space.
As a proof of concept, 
the authors present results using a linear function for support.
However, an open question is how to choose 
an appropriate support function
and its parameters.

\cite{Omiecinski2003} introduced several alternative
interest measures for associations which avoid the need for support
entirely. 
Two of the measures are {\em any-} and {\em all-confidence.}
Both rely only on the confidence measure defined for association rules.
Any-confidence is defined as the largest confidence of a rule 
which can be generated using all items from an itemset. The
author states that although finding all itemsets with a set
any-confidence would enable us to find all rules with a given
minimum confidence, any-confidence cannot be used efficiently as a
measure of interestingness since minimum confidence is not downward
closed. 
The all-confidence measure is defined as the smallest
confidence of all rules which can be produced from an set of
associated items. 
\citeauthor{Omiecinski2003} shows that a minimum constraint on 
all-confidence is
downward closed and, therefore, can be used for efficient mining
algorithms without support.

Another family of approaches is based on using statistical methods
to mine associations.  The main idea is to identify associations as
significant deviations from a baseline given by the assumption that
items occur statistically independent from each other. The
simplest measure to quantify this deviation is 
{\em interest}~\citep{Brin1997} which is often also called {\em lift}.
%
Interest for a rule $X \longrightarrow  Y$ is defined as
%
%
$P(X \cup Y)/(P(X)P(Y))$,
where the denominator is the baseline probability, the expected
probability of the itemset under independence.  Interest is usually
calculated by the ratio $r_{obs}/r_{exp}$ which are the observed
and the expected occurrence counts of the itemset.
%
The ratio is close to one if the itemsets $X$ and $Y$ occur 
together in the
database as expected under the assumption
that they are independent. A value greater than one indicates a
positive correlation between the itemsets and values lesser than one
indicate a negative correlation.  
To smooth away noise for low counts in the interest ratio, 
\cite{DuMouchel2001} developed
the empirical Bayes Gamma-Poisson shrinker. 
However, the interest ratio is not a
frequency constraint and does not possess the downward closure
property needed for efficient mining.

\cite{Silverstein1998} suggested mining
dependence rules using the $\chi^2$~test for independence between
items on $2 \times 2$ contingency tables.  The authors use the fact
that the test statistic can only increase with the number of items
to develop mining algorithms which rely on this 
{\em upward closure property.} 
\cite{DuMouchel2001} pointed out
that more important than the test statistic is the test's p-value.
Due to the increasing number of degrees of freedom of the
$\chi^2$~test the p-value can increase or decrease with itemset size, which 
invalidates the upward closure property.
Furthermore, \cite{Silverstein1998} mention that
a significant problem of the approach is the normal approximation
used in the $\chi^2$~test. This can skew results unpredictably for
contingency tables with cells with low expectation.

First steps towards the approach presented in this paper
were made with two projects  concerned with finding
related items for recommendation 
systems~\citep{GeyerSchulz2002b, GeyerSchulz2003c}.
The used algorithms were based on 
the logarithmic series distribution (LSD) model
which is a simplification of the NB model used in this paper.
Also the algorithms were restricted to find only $2$-itemsets.
However, the projects showed that the approach described in this paper
produces good results for real-world applications.

\section{Developing a Model-Based 
	Frequency Constraint\label{development}}
In this section we build on the idea of discovering associated
items with the help of observed deviations of co-occurrences from a
baseline which is based on independence between all items.
This is similar to how interest (lift) uses the expected probability
of itemsets under independence to identify dependent itemsets.
In contrast to lift and other similar measure, we
will not estimate the degree of deviation at the level of an
individual itemset. Rather, we will evaluate the deviation for
the set of all possible $1$-extensions of an itemset
together to find a local frequency constraint for these extensions. 
A $1$-extension
of an $k$ itemset is an itemset of size $k+1$ which is produced by
adding an additional item to the $k$-itemset.  

\subsection{A Simple Stochastic Baseline Model\label{modelNB}}
\begin{figure} \centering 
\includegraphics[scale=0.65]{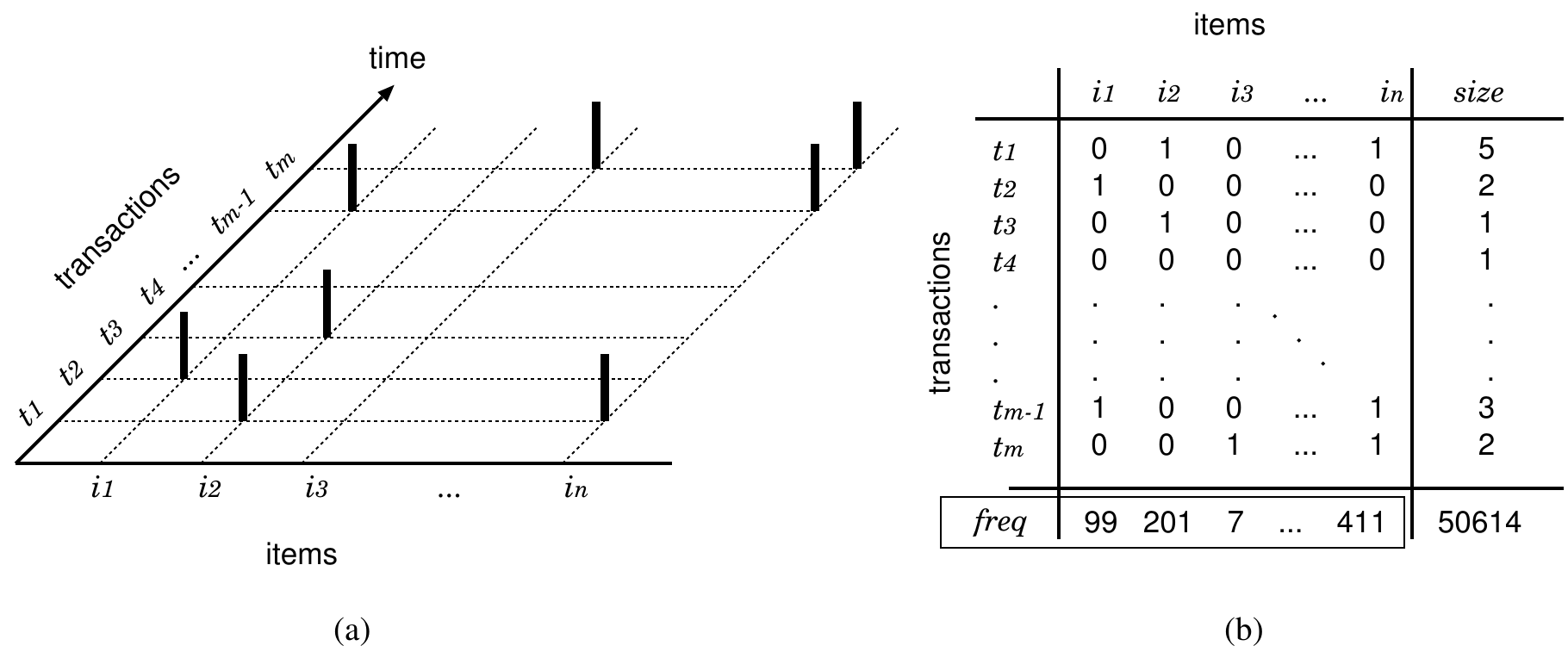}
\caption{Representation of an example database as (a) sequence of
transactions and (b) the incidence matrix.
\label{model}} 
\end{figure}

A suitable stochastic item occurrence model
for the baseline frequencies 
needs to describe the occurrence of independent
items with different usage frequencies
in a robust and mathematically tractable way.
For the model we consider the occurrence 
of items $I = \{i_1, i_2,\dots, i_n\}$ in a database 
with a fixed number of $m$ transactions.
An example database is depicted in Fig.~\ref{model}. For the
example we use $m=20,000$ transactions and $n=500$ items.  To the
left we see a graphical representation of the database as a
sequence of transactions over time.  The transactions contain items
depicted by the bars at the intersections of transactions and
items.  The typical representation used for data mining is the
$m \times n$ incidence matrix 
in Fig.~\ref{model}(b).  Each row sum represents the
size of a transaction and the column sums are the frequencies of
the items in the database.  The total sum represents the number of
incidences (item occurrences) in the database. Dividing the number
of incidences by the number of transactions gives the average
transaction size (for the example, $50,614/20,000=2.531$) 
and dividing the number
of incidences by the number of items gives the average item
frequency ($50,614/500=101.228$).

In the following we will model the baseline for the distribution
of the items' frequency counts $\textit{freq}$ in
Fig.~\ref{model}(b).
For the baseline we suppose that each item 
in the database follows an independent (homogeneous) Poisson process with an
individual latent rate $\lambda$. 
Therefore, the frequency for each item in the database is a value
drawn from the Poisson distribution with its latent rate.
We also assume that the individual rates are randomly drawn from a
suitable distribution defined by the continuous random variable
$\Lambda$.  Then the probability distribution of $R$, a random
variable which gives the number of times an arbitrarily chosen item
occurs in the database, is given by \nopagebreak

\begin{equation} 
Pr[R=r]=\int_0^\infty  \frac{e^{-\lambda}
\lambda^{r}}{r!} dG_\Lambda(\lambda),\;r=0,1,2,...,\;\lambda>0.
\label{mixture} 
\end{equation}

This Poisson mixture model results from the continuous
mixture of Poisson distributions with rates following the mixing
distribution $G_\Lambda$.

Heterogeneity in the occurrence frequencies between items is
accounted for by the form of the mixing distribution.
A commonly used and very flexible mixing distribution is the Gamma
distribution with the density function

\begin{equation} 
g_\Lambda(\lambda)=\frac{e^{-\lambda/a}
\lambda^{k-1}}{a^k\Gamma(k)},\;a>0,\;k>0,
\label{gamma} 
\end{equation}

where $a$ and $k$ are the distribution's 
scaling and the shape parameters.

Integrating Eq.~(\ref{mixture}) with (\ref{gamma}) 
is known to result in the negative binomial (NB)
distribution 
(see, e.g., \cite{Johnson1993}) 
with the probability distribution

\begin{equation}
Pr[R=r]=(1+a)^{-k}\frac{\Gamma(k+r)}{\Gamma(r+1)\Gamma(k)}
	\left(\frac{a}{1+a}\right)^{r},\;r=0,1,2,...  
\label{nb} 
\end{equation}

This distribution gives the probability that we see arbitrarily
chosen items with a frequency of $r=0,1,2,...$ in the database.
The average frequency of the items in the database is given by
$a/k$
and $Pr[R=0]$ represents the proportion of available
items which never occurred during the time the database was
recorded. 

Once the parameters $k$ and $a$ are known, the expected 
probabilities of finding items with a frequency of $r$ 
in the database
can be efficiently computed by calculating the probability of the
zero class by $Pr[R=0]= (1+a)^{-k}$ and then using the recursive
relationship (see~\cite{Johnson1993})

\begin{equation} 
Pr[R=r+1]= \frac{k+r}{r+1}\;\frac{a}{1+a} 
	\;Pr[R=r]. 
\label{nb_recursion} 
\end{equation}


%
Although, the NB model (often also called Gamma-Poisson model)
simplifies reality considerably with its assumed Poisson processes
and the Gamma mixing distribution, it is widely and successfully
applied for
accident statistics, birth-and-death processes, economics, library
circulation, market research, medicine, and military
applications~\citep{Johnson1993}.  

\subsection{Fitting the Model to Transaction Data
Sets\label{fitting}}

The parameters of the NB distribution can be estimated 
by several
methods including the method of moments, maximum likelihood, and
others~\citep{Johnson1993}.  
All methods need the 
item frequency counts $\textit{freq}$
for the estimation. This information is obtained by passing over
the database once. Since this counts are necessary to calculate
the item support needed by most mining algorithms, the overhead 
can be saved later on when itemsets are mined.

Particularly simple is the method of moments
where 
$\Tilde{k}= {\Bar{r}^2}/{(s^2-\Bar{r})}$ 
and
$\Tilde{a}=\Bar{r} / \Tilde{k}$ can be directly computed from the
observed mean 
$\Bar{r}=\mathrm{mean}(\textit{freq})$
and variance 
$s^2 = \mathrm{var}(\textit{freq})$
of the item occurrence
frequencies.  However, with empirical data we face two problems:
(a) 
the zero-class (available items which never occurred in the
database) are often not observable and
(b) 
as reported for other applications of the NB model, in real-world
data often exist a small number of
items with a too high frequency to be covered by the Gamma mixing
distribution used in the model. 

A way to obtain the missing zero-class is to subtract the number of
observed items from the total number of items which were available
at the time the database was recorded.  The number of available
items can be obtained from the provider of the database.  
If the total number of available items is unknown, the size of the
zero-class can be estimated together with the parameters of the
NB distribution.
The standard procedure for this type of estimation problem is the
{\em Expectation Maximization (EM)} algorithm~\citep{Dempster1977}.
This procedure 
iteratively estimates missing values using the observed data
and the model using intermediate values of the parameters, and then uses the
estimated data and the observed data to update the parameters for
the next iteration. The procedure stops when the parameters
stabilize. For our estimation problem 
the procedure is computationally very inexpensive.
Each iteration involves 
only to calculate
$n(1+\tilde{a})^{-\tilde{k}}$ to estimate 
the count for the missing zero-class and then 
applying the method of moments (see above) to update the parameter estimates 
$\tilde{a}$ and $\tilde{k}$.
As we will see in the examples later in this section, 
the EM algorithm usually only needs
a small number of iteration to estimate the needed parameters.
Therefore, the computational cost of estimation is 
insignificant compared to the 
time needed to count the item frequencies in the database.

The second estimation problem are outliers with too high frequencies.  These
outliers will distort the mean and the variance and thus will lead to a model
which grossly overestimates the probability of seeing items with high
frequencies.  For a more robust estimate, we can trim a suitable percentage of
the items with the highest frequencies. A suitable percentage can be found by
visual comparison of the empirical data and the estimated model or by
minimizing the $\chi^2$-value of the goodness-of-fit test.

To demonstrate that the parameters for the developed baseline 
model can be estimated for data sets, we use the two
e-commerce data sets {\em WebView-1} and {\em POS} provided by Blue
Martini Software for the KDD Cup 2000~\citep{kddcup2000} and an
artificial data set, {\em Artif-1}.  WebView-1 contains several
months of clickstream data from an e-commerce Web site where
each transaction consists of the product detail page views during a
session.  POS is a point-of-sale data set containing several years
of data.  Artif-1 is better known as {\em T10I4D100K}, a widely
used artificial data set generated by the procedure described 
by \cite{Agrawal1994}.

\begin{table}
\caption{Characteristics of the used data sets.\label{charcteristics}} 
\vspace{3mm}

\centering
\begin{tabular}{lrrrr} 
\hline  
	&{\bf WebView-1}& {\bf POS} & {\bf Artif-1} 
\\ 
\hline  
Transactions&59,602&515,597&100,000 \\
Avg. trans. size&2.5&6.5&10.1 \\
Median trans. size&1&4&10 \\
Distinct items&497&1,657&844 \\ 
\hline  
\end{tabular} 
\end{table}

Table~\ref{charcteristics} contains the basic characteristics of
the data sets. The data sets differ in the number of items and the
average number of items per transactions.  The real-world data sets
show that their median transaction size is considerably smaller
than their mean which indicates that the distribution of
transaction lengths is skewed with many very short transactions and
some very long transactions.  The artificial data set does not show
this property. 
For a comparison of the data sets' properties and their 
impact on the effectiveness of different association rule 
mining algorithms we refer to
\cite{Zheng2001}\footnote{Although the 
artificial data set in this paper and in 
\cite{Zheng2001} were produced using the 
same generator (available at
http://www.almaden.ibm.com/software/quest/Resources/), there are
minimal variations due to differences in the used 
random number generator initialization.}.

Before we estimated the model parameters with the EM~algorithm, 
we discarded the first 10,000 transactions
for WebView-1 since a preliminary data screening showed that the
average transaction size and the number of items used in these
transactions is more volatile and significantly smaller than for
the rest of the data set.  This might indicate that at the
beginning of the database there were still major changes made to
the Web shop (e.g., reorganizing the Web site, or adding and
removing promotional items).  POS and Artif-1 do 
not show such effects.
To remove outliers (items with too high frequencies), we used visual inspection
of the item frequency distributions and the fitted models for a range of
trimming values (between 0 and 10\%).  To estimate the parameters for the two
real-world data sets we chose to trim 2.5\% of the items with the highest
frequency. The synthetic data set does not contain outliers and therefore no
trimming was necessary.

\begin{table}
\caption{The fitted NB models using samples of
	20,000 transactions.  \label{estim_table}} 
\vspace{3mm}

\centering
\begin{tabular}{lrrrr} 
\hline  
	&\bf WebView-1 &\bf POS &\bf Artif-1\\ 
\hline  
Observed items & 342  &  1,153 & 843 \\
Trimmed items & 9 & 29 & 0 \\
Item occurrences & 33,802 & 87,864 & 202,325 \\ 
\hline
EM iterations &3&29&3\\
Estim. zero-class & 6 & 2,430 & 4 \\
Used items ($\Tilde{n}$) & 339 & 3,554 & 847 \\
$\Bar{r}$ & 99.711 & 24.723 & 238.873 \\
$s^2$ & 11,879.543 &  9,630.206 & 59,213.381 \\
$\Tilde{k}$ & 0.844 &   0.064 & 0.968 \\
$\Tilde{a}$ & 118.141 &   386.297 & 242.265 \\
\hline
$\chi^2$ p-value & 0.540 &  0.101 &  0.914 \\
\hline 
\end{tabular} 
\end{table}

In Table~\ref{estim_table}, we summarize the results of the fitting
procedure for samples of size 20,000 transactions from the three
data sets. To check whether the model provides a useful
approximation for the data, we used the $\chi^2$~goodness-of-fit
test.  As recommended for the test, we combined classes so that in
no class the expected count is below $5$ and used a statistical
package to calculate the p-values.  For all data sets we found
high p-values ($\gg 0.05$) which indicates that no significant
difference between the data and the corresponding models could 
be found and the model fits the data sets reasonably well.

To evaluate the stability of the model parameters, we 
estimated the parameters for samples of different sizes. 
We expect that the shape
parameter $k$ is independent of the sample size while the scale
parameter $a$ depends linearly on the sample size.  This can be
simply explained by the fact that, if we, observe the Poisson 
process for each item, e.g., twice as long, 
we have to double the latent parameter $\lambda$ 
for each process. For the Gamma mixing
distribution this means that the scale parameter $a$ must be
double.  Consequently, $a$ divided by the size of the sample
should be constant.

Table~\ref{stability} gives the parameter estimates (also $a$ per
transaction) and the estimated total number of items $n$ (observed
items + estimated zero class) for samples of sizes $10,000$ to
$40,000$ transactions from the three databases.  The estimates for
the parameters $k$, $a$ per transaction, and the number of items $n$
generally show only minor variations over different sample sizes 
of the same data set. We analyzed the reason for the 
high jump of the estimated 
number of items from $339$ for $20,000$ transactions to $395$
for $40,000$ transactions in WebView-1.
We found evidence in the database that
after the first $20,000$ transactions
the number of different items in the database starts to grow
by about $10$ items every $5,000$ transactions.
However, this fact does not
seem to influence the stability of the estimates of the 
parameters $k$ and $a$.
The stability enables us to use model
parameters estimated for one sample size for samples of 
different sizes.

\begin{table} 
\caption{Estimates for the NB-model using samples
	of different sizes.\label{stability} } 
\vspace{3mm}
\centering
\begin{tabular}{lrrrrr}
\hline 
Name	&Sample size & $ \Tilde{k}$ & $\Tilde{a}$
	&$\Tilde{a}$ per transaction  & $\Tilde{n}$ 
\\
\hline  
{\bf WebView-1}&	10,000& 0.933 & 58.274 & 0.0058	& 325 \\ 
{\bf WebView-1}&	20,000& 0.844 & 118.140& 0.0059 & 339 \\ 
{\bf WebView-1}&	40,000& 0.868 & 218.635& 0.0055	& 395 \\
\hline 
{\bf POS}&		10,000& 0.060 & 178.200& 0.0178 &3,666 \\
{\bf POS}& 		20,000& 0.064 & 386.300& 0.0193 &3,554 \\
{\bf POS}&		40,000& 0.064 & 651.406& 0.0163	&3,552 \\
\hline 
{\bf Artif-1}&		10,000& 0.975 & 123.313 &0.0123	&845 \\
{\bf Artif-1}&		20,000& 0.968 & 242.265&0.0121 &847  \\
{\bf Artif-1}&		40,000& 0.967 & 493.692&0.0123	&846 \\
\hline 
\end{tabular} 
\end{table}

\begin{figure} 
\begin{minipage}[b]{.48\linewidth}
\includegraphics[scale=0.48]{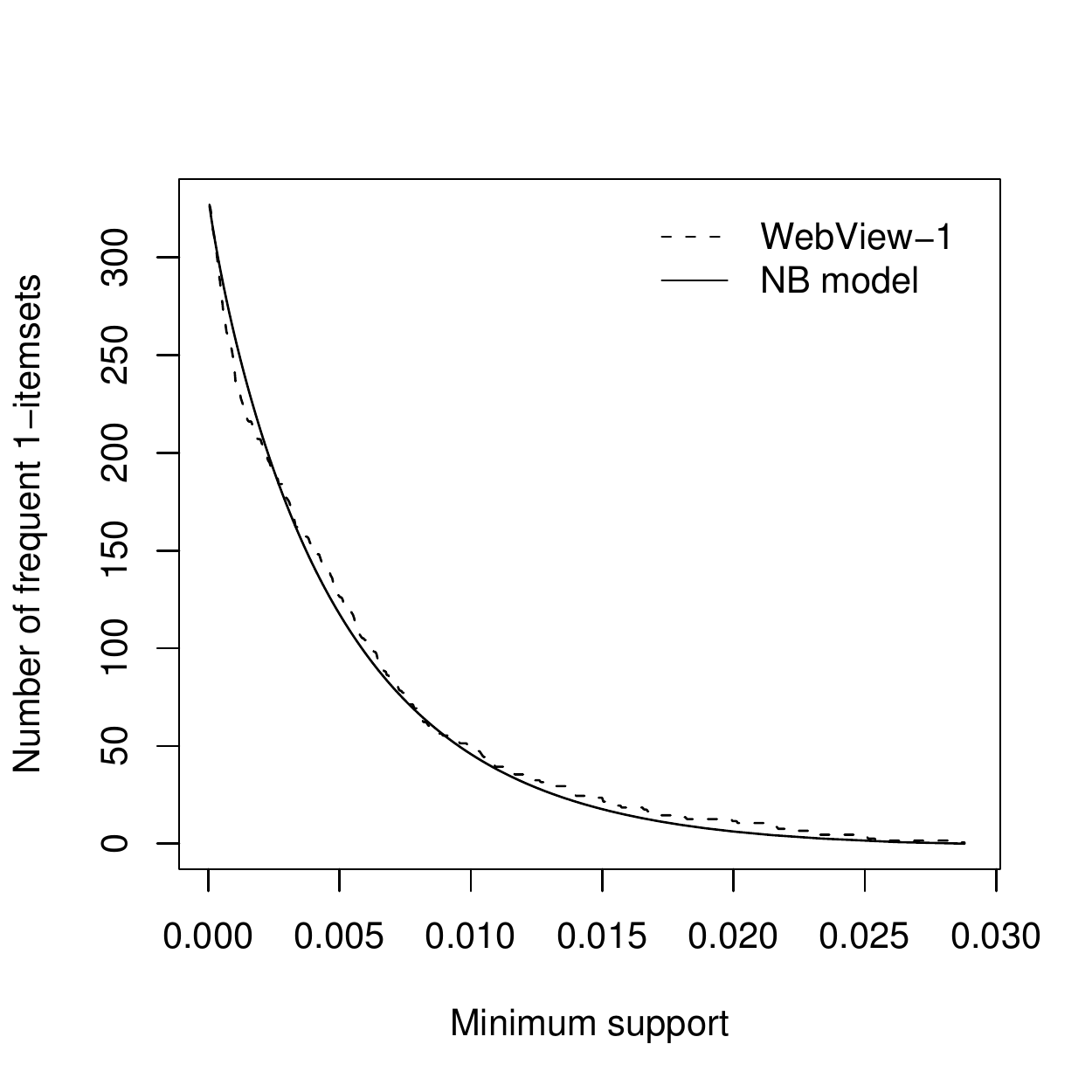}
\end{minipage}
\begin{minipage}[b]{.48\linewidth}
\includegraphics[scale=0.48]{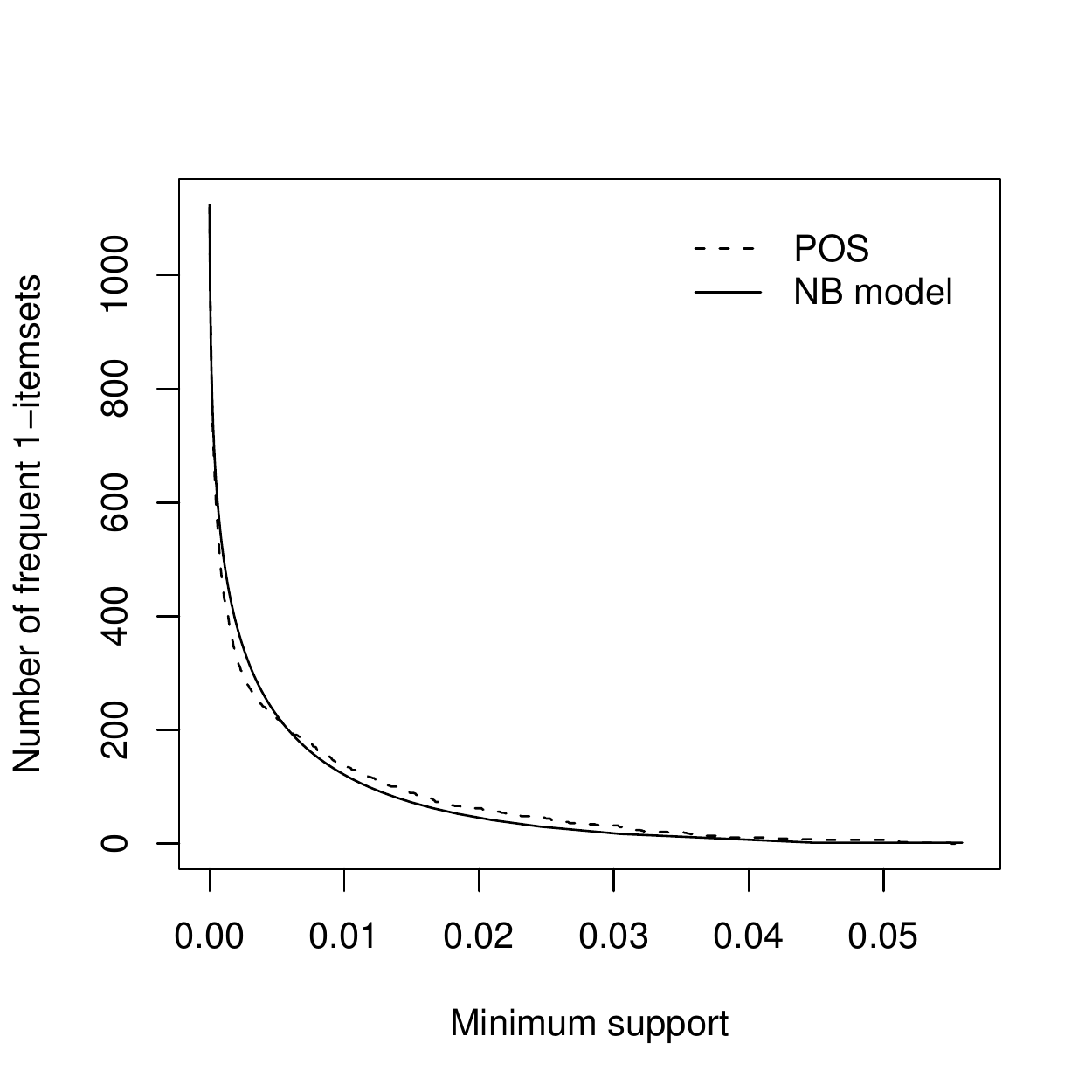}
\end{minipage}

\begin{minipage}[b]{.48\linewidth}
\includegraphics[scale=0.48]{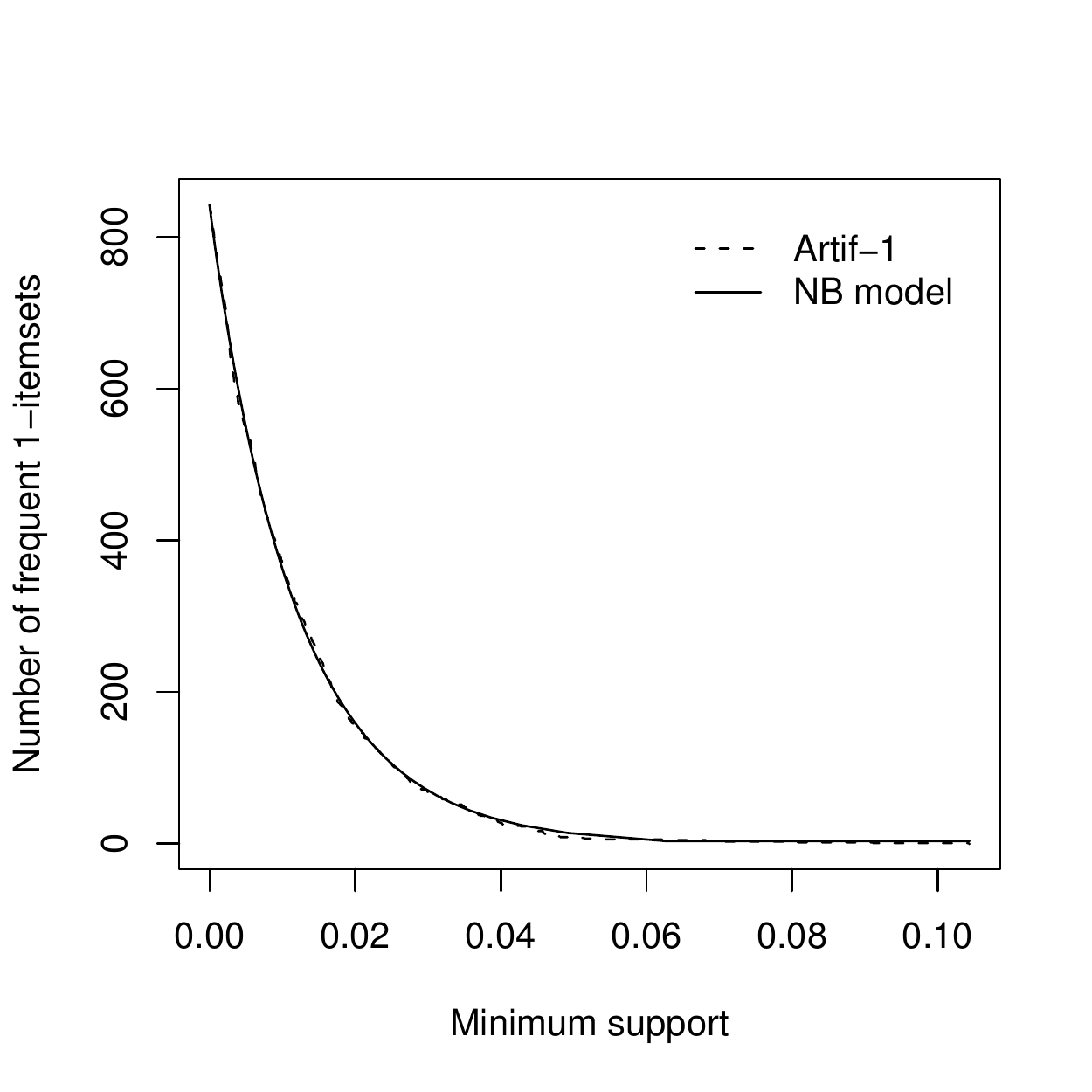}
\end{minipage}
\caption{Actual versus predicted number of frequent items by
	minimum support. \label{estim_fig}} 
\end{figure}

Applied to associations, Eq.~(\ref{nb}) in the section above gives the
probability distribution of observing single items ($1$-itemsets)
with a frequency of $r$.  
Let $\sigma^{\mathit{freq}} = \sigma m$, 
where $m$ is the number of transactions in the database,
be the frequency threshold equivalent to the minimum support $\sigma$. 
Then the expected number of
$1$-itemsets 
which satisfy the frequency threshold $\sigma^{\mathit{freq}}$ 
is given by 

\begin{displaymath} 
n  Pr[R \ge \sigma^{\mathit{freq}}], 
\end{displaymath} 

where $n$ is the number of available items.
In Fig.~\ref{estim_fig}
we show for the data sets 
the number of frequent
$1$-itemsets predicted by the fitted models (solid line) and the
actual number (dashed line) by a varying minimum support
constraint.  For easier comparison we show relative support for the
plots.  In all three plots we can see how the models fit the
skewed support distributions.

\subsection{Extending the Baseline Model to $k$-Itemsets
	\label{expand}}

After only considering $1$-itemsets, we show how the model
developed above can be extended to provide a baseline for the
distribution of support 
over all possible $1$-extensions of an itemset. 

\begin{figure} 
\centering 
\includegraphics[scale=0.7]{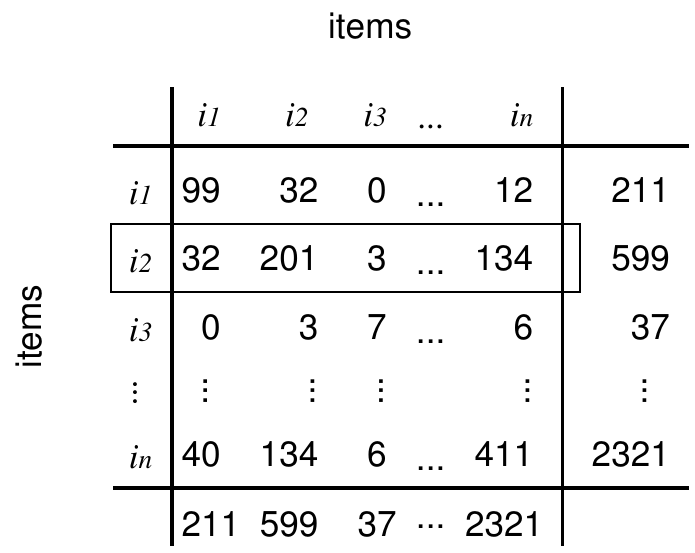}
\caption{A $n \times n$ matrix for counting $2$-itemsets in the
	database.\label{model2}} 
\end{figure}

We start with $2$-itemsets before we generalize to itemsets of
arbitrary length.  Fig.~\ref{model2} shows an example of the
co-occurrence frequencies of all items (occurrence of $2$-itemsets)
in transactions organized as an $n \times n$ matrix.  The matrix is
symmetric around the main diagonal which contains the count
frequencies of the individual items 
$\mathrm{freq}(i_1), \mathrm{freq}(i_2), \ldots, \mathrm{freq}(i_n)$.  
By adding the
count values for each row or for each column, we get 
in the margins of the matrix the number of incidences 
in all transactions which contain the respective item.

For example, to build the model for all $1$-extensions of item $i_2$, we only
need the information in the box in Fig.~\ref{model2}.
It contains the frequency counts for all $1$-extensions 
of $i_2$ plus $\mathrm{freq}(i_2)$ in cell $(2,2)$.  
Note, that these counts are
only affected by transactions which contain item $i_2$.
If we select all transactions which contain item
$i_2$, we get a sample of size $\mathrm{freq}(i_2)=201$ from the
database.
For the baseline model with only independent items,
the co-occurrence counts in the sample 
follow again Poisson processes. Following the model in
Section~\ref{modelNB} we can obtain a new random variable $R_{i_2}$
which models the occurrences of an arbitrarily chosen
$1$-extensions of $i_2$.  

After presenting the idea for the $1$-extensions of a single item, we 
now turn to the general case of building a baseline model for all
$1$-extensions of an association $l$ of arbitrary length.
We denote the number of items in $l$ by $k$. Thus $l$ is a
$k$-itemset for which exactly $n-k$ different $1$-extensions exist.
All $1$-extensions of $l$ can be generated by joining $l$
with all possible single items $c \in {I \setminus l}$.  
The items $c$  will be call candidate items.
In the baseline model all candidate items are independent
from the items in $l$. Consequently, the set of all
transactions which contain $l$ represent a sample of size
$\mathrm{freq}(l)$, which is random with respect to the candidate items.  
Following the developed model also the baseline for the number of candidate 
items with frequency $r$ in the sample 
has a NB distribution.
More precisely, 
the counts for the $1$-extensions of $l$ can be modeled by a random variable
$R_l$ with the probability distribution 

\begin{equation}
Pr[R_l=r]=(1+a_l)^{-k}\frac{\Gamma(k+r)}{\Gamma(r+1)\Gamma(k)}
	\left(\frac{a_l}{1+a_l}\right)^{r}\;for\ r=0,1,2,...  
\label{nbl} 
\end{equation}

The distribution's shape parameter $k$ is not affected by sample
size and we can use the estimate $\Tilde k$ 
from the database.  However, the
parameter $a$ is linearly dependent on the
sample size (see Section~\ref{fitting} above).
To obtain $a_l$, we have to rescale $\Tilde a$, 
estimated from the database,
for the sample size $\mathrm{freq}(l)$.

To rescale $a$ we could use the proportion of the transactions in
the sample relative to the size of the database which was used to
estimate $\Tilde a$.  In Section~\ref{fitting} above, we showed that
for estimating the parameter for different sample sizes gives a
stable value for $\Tilde a$ per transaction.  A problem with
applying transaction-based rescaling is that the more items we
include in $l$, the smaller the number of remaining items per
transaction gets. This would reduce the effective transaction
length and the estimated model would not be applicable.
Therefore, we will ignore the concept of transactions for the
following and treat the data set as a series of incidences 
(occurrences of items). 
For the baseline model this is
unproblematic since the mixture model never used the information
that items occur together in transactions.
At the level of incidences, we can rescale $a$ by the proportion of incidences in the
sample relative to the total number of incidences in the database
from which we estimated the parameter.  
We do this in two steps:

\begin{enumerate} 
\item 
We calculate $\Tilde a'$, the parameter per
incidence, by dividing the parameter obtained from the database by
the total number of incidences in the database.

\begin{equation} 
\Tilde a'= \frac {\Tilde a} { \sum_{t \in {\cal D}} |t|} 
\label{nb_scalea} 
\end{equation}

\item We rescale the parameter for itemset
$l$ by multiplying $\Tilde a'$ with the number of incidences in the sample (transactions which contain $l$) excluding 
the occurrences of the items in $l$.

\begin{equation} 
\Tilde a_l=\Tilde a'  \sum_{\{t \in {\cal D} | t \supset l\}} |t \setminus l| 
\label{nb_scalea2}
\end{equation}
%
\end{enumerate}

For item $i_2$ in the example in Fig.~\ref{model2}, the rescaled
parameter can be easily calculated from the sum of incidences for
the item ($599$) in the $n \times n$ matrix together with the the
sum of incidences ($50,614$) in the total incidence matrix (see
Fig.~\ref{model} above in Section~\ref{modelNB}) 
by $\Tilde a' = \Tilde a / 50614$ and 
$\Tilde a_{i_2} = \Tilde a' \cdot 599$.

\subsection{Deriving a Model-Based Frequency Constraint for
NB-Frequent Itemsets\label{constraint}}

The NB distribution with the parameters rescaled for itemset $l$
provides a baseline for the frequency distribution of the candidate
items in the transactions which contain $l$, i.e., 
the number of different
itemsets $l \cup \{c\}$ with $c \in I \setminus l$ we would expect
per support count, if all items were independent.  If in the database some
item candidates are related to the items in $l$, the transactions
that contain $l$ cannot be considered a random sample for these
items. These related items will have a higher frequency
in the sample than expected by the baseline model. 

To find a set $L$ of non-random $1$-extensions of $l$ 
(extensions with item candidates with a too high co-occurrence frequency), 
we need to identify a
frequency threshold $\sigma^{\mathit{freq}}_l$, where accepting
item candidates with a frequency count 
$r \ge \sigma^{\mathit{freq}}_l$ 
separates associated items best from items which co-occur
often by pure chance.  For this task we need to define a quality
measure on $L$, the set of accepted $1$-extensions.  
{\em Precision} is a possible quality measure 
which is widely used 
by the machine
learning community~\citep{Kohavi1988}
and is defined as the proportion of correctly predicted positive cases
in all predicted positive cases.
Using the baseline model
and observed data, we can predict precision for different values of
the frequency threshold.

\begin{Def}[Predicted precision] 
Let $L$ be the set of all $1$-extensions
of a known association $l$ which are generated by joining 
$l$ with all
candidate items $c \in I \setminus l$ which co-occurrence with
$l$ in at least $\rho$ transactions. 
For set $L$ we
define the predicted precision as

%
\begin{equation}
\mathrm{precision}_l(\rho) = 
\begin{cases}
(o_{[r \ge \rho]} - e_{[r \ge \rho]}) / o_{[r \ge \rho]}  
& \text{if $o_{[r \ge \rho]} \ge e_{[r \ge \rho]}$ and $o_{[r \ge \rho]} > 0$} \\
0 & \text{otherwise.}
\label{prec} 
\end{cases}
\end{equation}

$o_{[r \ge \rho]}$
is the observed and $e_{[r \ge \rho]}$ is the expected 
number of candidate items which
have a co-occurrence frequency with itemset $l$ of $r \ge \rho$.
The observed number is calculated as the sum of
observations with count $r$ by
$o_{[r \ge \rho]}= \sum_{r=\rho}^{r_{max}} o_r$,
where $r_{max}$ is the highest observed co-occurrence.
The expected number is given by the baseline model
as 
$e_{[r \ge \rho]} = (n -|l|) Pr[R_l \ge \rho]$,
where $n-|l|$ is the number of possible
candidate items for pattern $l$.
\end{Def}

Predicted precision together with a precision threshold $\pi$ 
can be used to form a model-based constraint on accepted associations.
The smallest possible frequency threshold for
$1$-extensions of $l$, which satisfies the set
minimum precision threshold $\pi$, can be found by 

\begin{equation} \label{choose_sigma}
\sigma^{\mathit{freq}}_l =\mathrm{argmin}_\rho\{\mathrm{precision}_l(\rho) \ge \pi\}.
\end{equation}

The set of the chosen candidate items for $l$ is then

\begin{displaymath} 
C_l= \{c \in I \setminus l | 
\mathrm{freq}(l \cup \{c\}) \ge \sigma^{\mathit{freq}}_l\},
\end{displaymath} 

and the set of accepted associations is 

\begin{displaymath} 
L = \{l \cup \{c\} | c \in C_l\}.  
\end{displaymath}

The predicted error rate for using a threshold $\sigma_l^{\mathit{freq}}$ 
is given by $1-\mathrm{precision}_l(\sigma_l^{\mathit{freq}})$.
A suitable selection criterion
for a count threshold is to allow only a percentage of falsely
accepted associations. For example, if we need for an application all
rules with the antecedent $l$ and a single item as the consequent
and the maximum number of acceptable
spurious rules is 5\%, we can find all $1$-extension of
$l$ and use a minimum precision threshold
of $\pi = 0.95$.

Table~\ref{example1} contains an example for the model-based
frequency constraint using data from the WebView-1 database.  We
analyze the $1$-extensions of itemset $l=\{10311,12571,12575\}$ at
a minimum precision threshold of 95\%.  The estimates for $n$, $k$
and $a$ are taken from Table~\ref{estim_table} in
Section~\ref{fitting}.  Parameter $a$ is rescaled to
$a_l=1.164$ using Eqs.~(\ref{nb_scalea}) and (\ref{nb_scalea2}) 
in the previous section.
Column $o$ contains the observed number of items with a
co-occurrence frequency of $r$ with $l$. The value at
$r=0$ is in parentheses since it is not directly observable. It was
calculated as the difference between the estimated 
number of
available candidate items
($n - |l|$) and the number of observed items ($o_{[r>0]}$).
Column $e$ contains the
expected frequencies calculated with the model.  
To find
the frequency threshold $\sigma^{\mathit{freq}}_l$, the precision function
$\mathrm{precision}_l(\rho)$ in Eq.~(\ref{prec}) is evaluated starting 
with $\rho = r_{max}$ 
($18$ in the example in Table~\ref{example1}) and
$\rho$ is reduced till we get a predicted precision value which is below the
minimum precision threshold of $\pi=0.95$. 
The found frequency threshold is then
the last value for $r$, which produced a precision above the
threshold (in the example at $r=11$).
After the threshold is found, there is no need to evaluate the rest
of the precision function with $r<10$.  All candidate items with a
co-occurrence frequency greater than the found threshold are
selected.  For the example in Table~\ref{example1}, this gives 
a set of 6 chosen candidate items.

\begin{table}
\caption{An example for finding the frequency threshold at
$\pi=0.95$ (found at $r=11$).\label{example1}} 
\vspace{3mm}
\centering 
\begin{tabular}{cccc}
\hline  $r$ & $o$ & $e$ & $\mathrm{precision}(r)$
\\ \hline  
0 &  (183)  &176.71178 & - \\ 
1 &   81  & 80.21957 & - \\ 
2 & 48  & 39.78173 &- \\ 
3 &   13  & 20.28450 &- \\ 
4 &    6 & 10.48480 &- \\ 
5 &    0  &  5.46345 &- \\ 
6 &    1  & 2.86219 &- \\ 
7 &    0  &  1.50516 &- \\ 
8 &    1  & 0.79378 &- \\ 
9 &    0  & 0.41955 &- \\ 
10&     0 & 0.22214& 0.92108  \\ 
\fbox{11} &     2 & 0.11779& \fbox{0.95811} \\ 
12& 1 &   0.06253& 0.96661 \\ 
13&     1 & 0.03323& 0.97632 \\ 
14&     1 &   0.01767& 0.98109 \\ 
15&     0 & 0.00941& 0.97986 \\ 
16&     0 &   0.00501& 0.98927 \\ 
17&     0 & 0.00267& 0.99428 \\ 
18&     1 &   0.00305& 0.99695 \\
\hline  
\end{tabular} 
\end{table}

There exists an
interesting connection to the confidence measure
for the way an individual frequency threshold (minimum support) 
is chosen for all 
$1$-extensions of an itemset.

\begin{The}\label{equality}
Let $l$ be an itemset and let $c \in I \setminus l$
be the set of candidate items which form together with $l$ all
$1$-extensions of $l$.
For each possible minimum support $\sigma_l$ on the
$1$-extensions of $l$, 
a minimum confidence threshold
$\gamma_l$ 
on the rules $l\longrightarrow \{c\}$
exists, which results in an equivalent constraint. 
That is, there always exist pairs of values for $\sigma_l$
and $\gamma_l$ were the following holds:
\begin{displaymath} 
\mathrm{supp}(l \cup \{c\}) \ge \sigma_l \Leftrightarrow \mathrm{conf}(l\longrightarrow \{c\}) \ge \gamma_l.
\end{displaymath}
\end{The}

\begin{Pro} 
With $\mathrm{conf}(l\longrightarrow \{c\})$ defined as
$\mathrm{supp}(l \cup \{c\}) / \mathrm{supp}(l)$ we can
rewrite the right-hand side constraint as 
$\mathrm{supp}(l \cup \{c\})/\mathrm{supp}(l) \ge \gamma_l$. 
Since $\mathrm{supp}(l)$ is a positive constant for all considered rules,
we get the equality 
$\gamma_l = \sigma_l /\mathrm{supp}(l)$ by substitution.
\hfill $\Box$ 
\end{Pro}

As an example, suppose a database contains $20,000$ transactions and
the analyzed itemset $l$ is contained in $1600$ transactions which gives
$supp(l) = 1600/20,000 = 0.08$. 
If we require the candidate items $c$
to have a co-occurrence frequency with $l$ of at least 
$\mathrm{freq}(l \cup \{c\}) \ge 1200$, we use in fact
a minimum support of $\sigma_l = 1200/20,000 = 0.06$.
All rules $l \longrightarrow \{c\}$ which can be constructed for the
supported itemsets $l \cup \{c\}$ will have at least a confidence of 
$\gamma_l = 0.06 / 0.08 = 0.75$.

The aim of developing the model-based frequency constraint is to find as many
non-spurious associations as possible in a data base, given a precision
threshold.  After we introduced the model-based frequency constraint for
$1$-extensions of a single itemset, we now extend the view to the whole itemset
lattice.  For this purpose, we need to find a suitable search strategy 
which enables us to
traverse the itemset lattice efficiently, i.e. to prune parts 
of the search space which only contain itemsets which are not of interest.  
For frequent itemset mining, the downward closure property of support is 
exploited for this purpose. Unfortunately, the model-based frequency 
constraint does not possess such a
property. However, we can develop several search strategies.
A straight forward solution is to use an apriori-like 
level-wise search strategy (starting with 1-itemsets) and in every level~$k$ 
we only expand itemsets which passed the 
frequency constraint at level~$k-1$.
This strategy suffers from a problem with candidate items which are 
extremely frequent in the data base. For such a candidate item, we will 
always observe a high co-occurrence count with any, even unrelated itemsets.
The result is that itemsets which include a frequent but unrelated
item are likely to be used in the next level of the algorithm and 
possibly will be expanded even
further. In transaction data bases with a very skewed item frequency 
distribution this leads to many spurious associations and
combinatorial explosion.

Alternatively, since each $k$-itemset can be produced from $k$ 
different $(k-1)$-subsets (checked at level~$k-1$) plus the corresponding 
candidate item, it is also possible to require that for all $(k-1)$-subsets the 
corresponding candidate item passes the frequency constraint. This strategy
makes intuitively sense since for associated items one expects that each item
in the set is associated with the rest of the itemsets and thus should 
pass the constraint.
It also solves the problem with extremely frequent candidate items since it
is very unlikely that all unrelated and less frequent items pass by chance the
potentially high frequency constraint for the extremely frequent item.
Furthermore, this strategy prunes the search space significantly since an
itemset is only used for expansion if all subsets passed the frequency
constraint. However, the strategy has a problem with including
a relatively infrequent item into a set consisting of more frequent items. 
It is less likely that the infrequent item as the candidate item meets 
the frequency constraint set by the more frequent itemset, even if it is
related. Therefore it is possible that itemsets consisting of related 
items with varying frequencies are missed.

A third solution is to used a trade-off between the problems and pruning
effects of the two search strategies by requiring for a fraction $\theta$
(between one and all) of the subsets with their candidate items to pass the
frequency constraint. 
We now formally introduce the concept of 
{\em NB-frequent itemsets} which can be used to implement all three solutions:

\begin{Def}[NB-frequent itemset] 
A $k$-itemset $l'$ with $k>1$ is a NB-frequent itemset if, and only if, 
at least a fraction $\theta$
(at least one) of its $(k-1)$-subsets
$l \in \{l' \setminus \{c\} | c \in l'\}$ 
are NB-frequent itemsets and satisfy
$\mathrm{freq}(l \cup \{c\}) \ge \sigma^{\mathit{freq}}_l$.  The
frequency thresholds $\sigma_l^{\mathit{freq}}$ 
are individually chosen for each itemset $l$ using
Eq.~(\ref{choose_sigma}) with a user-specified 
precision threshold $\pi$.
All itemsets of size $1$ are per definition NB-frequent.  
\end{Def}

This definition clearly shows that NB-frequency in general is not
downward closed since only a fraction $\theta$ of the
$(k-1)$-subsets of a NB-frequent set of size $k$ are required to be
also NB-frequent.  Only the special case with $\theta=1$ offers
downward closure, but since the definition of NB-frequency is
recursive, we can only determine if an itemset is NB-frequent if we
first evaluate all its subsets.  However, the definition enables us to
build algorithms which find all NB-frequent itemsets in a bottom-up
search (expanding from 1-itemsets) and even to prune the search
space. The magnitude of pruning depends on the setting for parameter
$\theta$. 

Conceptually, mining NB-frequent itemsets with the extreme values
$0$ and $1$ for $\theta$ is similar to using 
Omiecinski's~\citeyearpar{Omiecinski2003}
any-con\-fi\-dence and all-confidence.
In
Theorem~\ref{equality} we showed that the minimum support
$\sigma_l$ chosen for NB-frequent itemsets $l \cup \{c\}$ is 
equivalent to choosing a minimum on confidence 
$\gamma_l = \sigma_l / \mathrm{supp}(l)$ 
for the rules $l \longrightarrow \{c\}$.  An
itemset passes a threshold on any-confidence if at least one rule
can be constructed from the itemset which has a confidence value
greater or equal of the threshold.  This is similar to mining
NB-frequent itemsets with $\theta=0$, where 
to accept itemset $l \cup \{c\}$ a single combination
$\mathrm{conf}(l \longrightarrow \{c\}) \ge \gamma_l$ 
suffices.

For all-confidence, all rules which can be constructed from an
itemset must have a confidence greater or equal than a threshold.
This is similar to mining NB-frequent itemsets with $\theta=1$
where we require 
$\mathrm{conf}(l \longrightarrow \{c\}) \ge \gamma_l$
for all possible combination. 
Note, that in contrast to all-
and any-confidence, we do not use a single threshold for mining
NB-frequent itemsets, but an individual threshold is chosen by the
model for each itemset $l$.

\section{A Mining Algorithm for NB-Frequent Itemsets
	\label{algorithms}}

In this section we develop an algorithm using a depth-first search
strategy to mine all NB-frequent itemset in a database.  
The algorithm implements the candidate item selection mechanism of
the model-based frequency constraint in the  NB-Select function.
The function's pseudocode is presented in Table~\ref{algo_NB-Select}. 
It is called for each found association $l$ and 
gets count information of all $1$-extensions of $l$,
characteristics of the data set ($n$, $\Tilde k$, $\Tilde a'$), and the user-specified precision
threshold $\pi$. 
NB-Select returns the set of selected candidate items for $l$.  
%

\begin{table} 
\caption{Pseudocode for selecting candidate items for the NB-frequent 
itemset $l$
using a minimum precision constraint.  \label{algo_NB-Select}}
\vspace{3mm}
\centering 
\begin{tabular}{c} 
\hline 
\begin{minipage}{0.9\textwidth}
\vspace{1mm}
{\bf function NB-Select}$(l, {\cal L}, n, \Tilde k, \Tilde a', \pi)$:
\begin{description}
\item
[$l$] = the itemset for which candidate items are selected
\item
[${\cal L}$] = a data structure which holds all
candidate items $c$ together with the associated counts
$c.count$
\item
[$n$] = the total number of available items in the database
\item
[$\Tilde k, \Tilde a'$] = estimated parameters for the database
\item
[$\pi$] = user-specified precision threshold
\end{description}
%
1. $r_\mathit{max} = max\{c.count | c \in {\cal L}\}$
\\2. $\mathit{rescale} = sum\{c.count | c \in {\cal L}\}$
\\3. {\bf foreach} count $c.count \in {\cal L}$ {\bf do}
$o_{[c.count]}\mathit{++}$
\\ 4. $\rho = r_{max}$
\\ 5. {\bf do}
\\6. \hspace{1.5mm} $\mathrm{precision} = 1 -
(n-|l|) Pr[R_l \ge \rho | k=\Tilde{k},
a = \Tilde{a}' \mathit{rescale} ]/\sum_{r=\rho}^{r_\mathit{max}}o_r$
\\7. {\bf while} ($\mathrm{precision} \ge \pi \wedge \rho\mathit{--} > 0$)
\\8. $\sigma^\mathrm{freq} = \rho+1$
\\9. {\bf return} $\{c \in {\cal L} | c.count \ge \sigma^\mathrm{freq} \}$
\vspace{1mm}
\end{minipage} 
\\
\hline  
\end{tabular}
\end{table}

Table~\ref{algo_DFS} contains the pseudocode for NB-DFS, the main
part of the mining algorithm.  The algorithm uses a similar 
structure as 
{\em DepthProject,} an algorithm to efficiently find long maximal
itemsets~\citep{Agarwal2000}.
NB-DFS is started with 
NB-DFS$(\emptyset,{\cal D},n, \Tilde k, \Tilde a', \pi, \theta)$ 
and recursively calls itself with 
the next analyzed itemset $l$ and its conditional database ${\cal D}_l$ 
to mine for subsequent
NB-frequent supersets of $l$. 
The conditional database ${\cal D}_l$ is a
sub-database which only contains transactions which contain $l$.
NB-DFS  scans  all transactions in the conditional database
to create the data structure ${\cal L}$ which stores the count 
information for the candidate items
and is needed by NB-Select.
%
New NB-frequent itemsets are generated with the NB-Gen function
which will be discussed later.
The algorithm stops when all NB-frequent itemsets are found.

\begin{table}[!t] 
\caption{Pseudocode for a recursive depth-first search algorithm
for NB-frequent itemsets.  \label{algo_DFS}} 
\vspace{3mm}
\centering 
\begin{tabular}{c} 
\hline 
\begin{minipage}{0.9\textwidth}
\vspace{1mm}
{\bf algorithm NB-DFS}$(l,{\cal D}_l, n, \Tilde k, \Tilde a', \pi, \theta)$:
\begin{description}
\item [$l$] = a NB-frequent itemset 
\item [${\cal D}_l$] = a conditional database only containing transactions which
include $l$
\item [$n$] = the number of all available items in the database
\item [$\Tilde k$, $\Tilde a'$] = estimated parameters for the database
\item [$\pi$] = user-specified precision threshold
\item [$\theta$] = user-specified required fraction of NB-frequent subsets
\item [${\cal L}$] = data structure for co-occurrence counts
\end{description}
1. ${\cal L} = \emptyset$
\\2. {\bf foreach} transaction $t \in {\cal D}_l$ {\bf do begin}
\\3. \hspace{3mm}
{\bf foreach} candidate item $c \in t \setminus l$ {\bf do begin}
\\4. \hspace{3mm}\hspace{3mm}
{\bf if} $c \in {\cal L}$ {\bf then} $c.count\mathit{++}$
\\5. \hspace{3mm}\hspace{3mm}\hspace{3mm}
{\bf else} add new counter $c.count=1$ to ${\cal L}$ 
\\6. \hspace{3mm} {\bf end}
\\7. {\bf end}
\\8. {\bf if} $l \ne \emptyset$ {\bf then} selected candidates
$C = $ NB-Select($l, {\cal L}, n, \Tilde k, \Tilde a', \pi$)
\\9. \hspace{3mm} {\bf else}
initial run candidates are $C=\{c \in {\cal L}\}$
\\10. delete or save data structure ${\cal L}$
\\11. $L= \text{NB-Gen}(l,C,\theta)$
\\12.
{\bf foreach} new NB-frequent itemset $l' \in L$ {\bf do begin}
\\13.\hspace{3mm} 
${\cal D}_{l'} =  \{t \in {\cal D}_l | t \supseteq l'\}$
\\14.\hspace{3mm} $L = L\ \cup$ NB-DFS($l', {\cal D}_{l'}, n, \Tilde k, \Tilde a', \pi, \theta$)
\\15. {\bf end}
\\16. {\bf return}  $L$
\vspace{1mm}
\end{minipage}
\\ \hline  
\end{tabular}
\end{table}

Compared to a level-wise breadth-first search algorithm, e.g. Apriori,
the depth-first algorithm uses significantly more passes over the
database.  However, every time only a conditional database is
scanned. This conditional database only contains the transactions
that include the itemset which is currently expanded.  
Note, that this conditional database contains all information
needed to find all NB-frequent supersets of the expanded itemset.
As this
itemset grows longer, the conditional database gets quickly smaller.
If the original database is too large to fit into
main memory, a conditional databases will fit into the memory after the
expanded itemset grew in size. This will make the subsequent scans very fast.

\begin{table} 
\caption{Pseudocode for the generation
function for NB-frequent itemsets.\label{NB-Gen}} 
\vspace{3mm}
\centering 
\begin{tabular}{c} 
\hline 
\begin{minipage}{0.9\textwidth} 
\vspace{1mm} 
{\bf function NB-Gen}$(l,C,\theta)$:
\begin{description}
\item [$l$] = a NB-frequent itemset
\item [$C$] = the set of candidate items chosen by NB-Select
for $l$
\item [$\theta$] = a user-specified parameter
\item [${\cal R}$] = a global repository
containing for each traversed itemset $l'$ of size $k$ an
entry $l'.\mathit{frequent}$ which is $\mathit{true}$
if $l'$ was already determined to be NB-frequent, and
a counter $l'.count$ to keep track of the 
number of NB-frequent $(k-1)$-subsets for which $l'$
was already accepted as a candidate.
\end{description}
1. $L =  \{l \cup \{c\} | c \in C\}$
\\ 2. {\bf foreach} candidate itemset $l' \in L$ {\bf do begin}
\\ 3. \hspace{3mm} {\bf if} $l' \notin {\cal R}$ {\bf then}
add $l'$ with $l'.\mathit{frequent}=\mathit{false}$
and $l'.\mathit{count}=0$ to ${\cal R}$
\\ 4. \hspace{3mm} {\bf if} $l'.\mathit{frequent} == true$  {\bf then}
delete $l'$ from $L$
\\ 5. \hspace{3mm} {\bf else} {\bf begin}
\\ 6. \hspace{6mm} $l'.count\mathit{++}$
\\ 7. \hspace{6mm} {\bf if} $l'.count < \theta |l'|$ {\bf then}
delete $l'$ from $L$
\\ 8. \hspace{9mm} {\bf else} $l'.\mathit{frequent} = \mathit{true}$
\\ 9. \hspace{3mm} {\bf end}
\\ 10. {\bf end}
\\ 11. {\bf return} $L$
\vspace{1mm} 
\end{minipage} 
\\ \hline 
\end{tabular} 
\end{table}

The generation function NB-Gen 
has a similar purpose as candidate generation in support-based algorithms: 
It controls what parts of the search space are pruned.  
Therefore, a suitable
candidate generation strategy is crucial for the performance of the mining
algorithm.  As already discussed, NB-frequency does not possess the downward
closure property which would allow pruning in the same way as for minimum
support.  
However, the definition of NB-frequent itemsets provides us with a way to
prune the search space.  From the definition we know that in order for a
$k$-itemset to be NB-frequent at least a proportion $\theta$ of its
$(k-1)$-subset have to be NB-frequent and produce the itemset together
with an accepted candidate item.  
Since for each $k$-itemset exist $k$ different subsets of size $k-1$, 
we only need to continue
the depth-first search for the $k$-itemset, for which we already 
found at least $k \theta$ NB-frequent
$(k-1)$-subset. This has a pruning effect on the search space size.

We present the pseudocode for the
generation functions in Table~\ref{NB-Gen}.  
The function is called for each found NB-frequent itemset $l$
individually and gets the set of accepted candidate items and the
parameter $\theta$.  
To enforce $\theta$ for the generation of a new NB-frequent
itemset $l'$ of size $k$, we need the information of how many
different NB-frequent subsets of size $k-1$ also produce $l'$.
And, at the same time, we need to make
sure that no part of the lattice is traversed more than once.
Other depth-first mining algorithms (e.g., FP-Growth or
DepthProject) solve this problem by using special representations
of the database (frequent pattern tree structures~\citep{Han2004} or
a lexicographic tree~\citep{Agarwal2000}).
These representations ensure that no part
of the search space can be traversed more than once.  However, these
techniques only work for frequent
itemsets using the downward closed minimum support constraint.  
To enforce
the fraction $\theta$ for NB-frequent itemsets and to ensure that
itemsets in the lattice are only traversed once by NB-DFS,
we use a global repository ${\cal R}$.
This repository  is used to keep track of the number
of times a candidate itemset was already generated and of the
itemsets which were already traversed.
This solution was inspired by the implementation of closed and 
maximal itemset filtering implemented
for the {\em Eclat} algorithm by~\cite{Borgelt2003}.

\section{Experimental Results\label{evaluation}}

In this section we analyze the properties and the effectiveness of
mining NB-frequent itemsets.  
%
To compare the performance of NB-frequent itemsets with existing
methods we use frequent itemsets and itemsets generated using
all-confidence as benchmarks.  We chose frequent itemsets since a
single support value represents the standard in mining association
rules.  All-confidence was chosen because of its promising properties
and its conceptual similarity with mining NB-frequent itemsets with
$\theta=1$.

\subsection{Investigation of the Itemset Generation Behavior}

First, we examine how the number of the NB-frequent itemsets found by the
model-based algorithm depends 
on the parameter $\theta$, which controls the magnitude of
pruning, and on the precision parameter $\pi$.  
For the generation function we use the settings with no and with maximal
pruning ($\theta=0$, $\theta=1$) and the intermediate value $\theta=0.5$ which
reduces the problems with itemsets containing items with extremely
different frequencies
(see discussion in section~\ref{constraint}).
Generally, we vary the
parameter $\pi$ for NB-Select between 0.5 and 0.999.  However,
since combinatorial explosion limits the range of practicable
settings, depending on the data set and the parameter $\theta$,
some values of $\pi$ are omitted.

\begin{figure} 
\begin{minipage}[b]{.48\linewidth}
\includegraphics[scale=0.48]{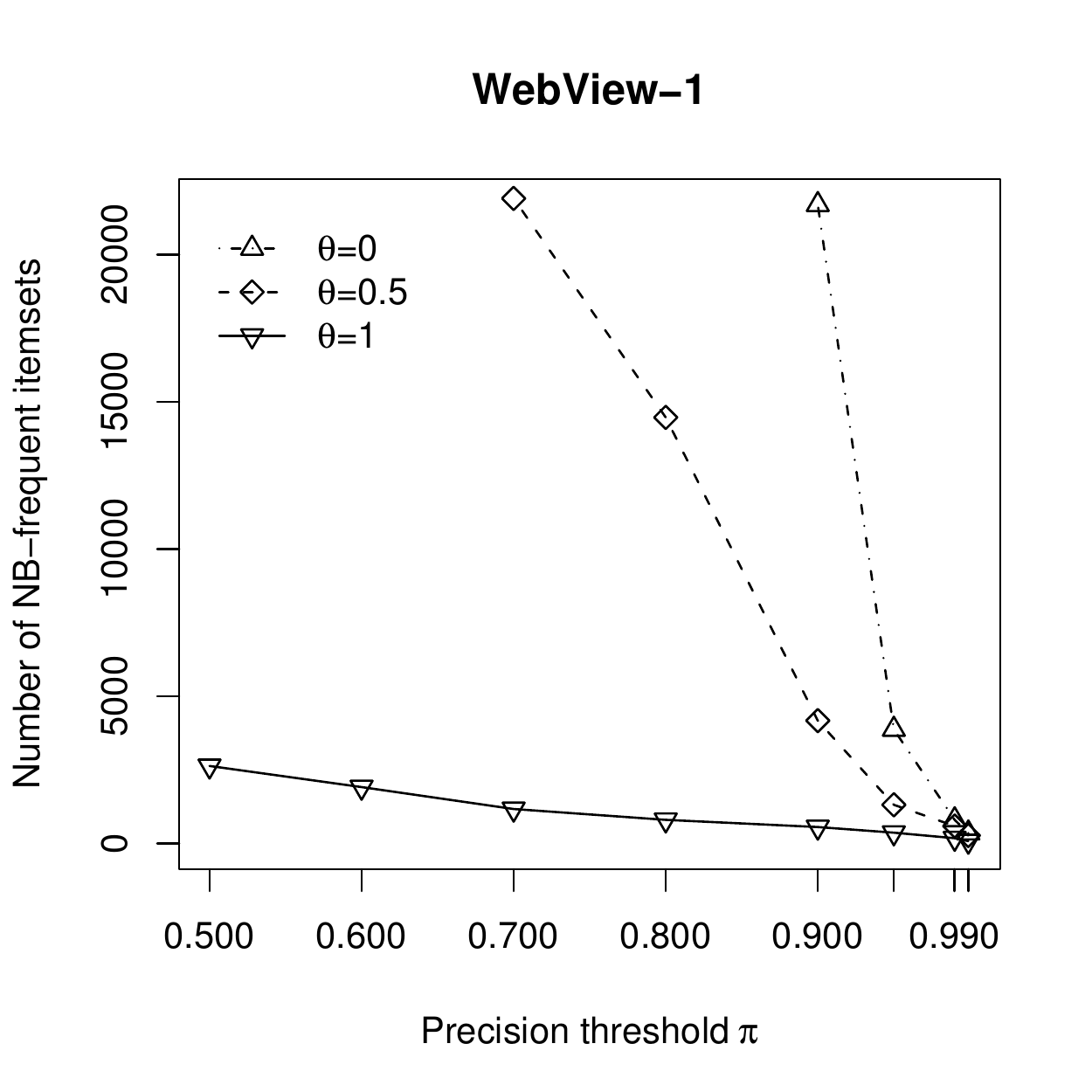}
\end{minipage}
\begin{minipage}[b]{.48\linewidth}
\includegraphics[scale=0.48]{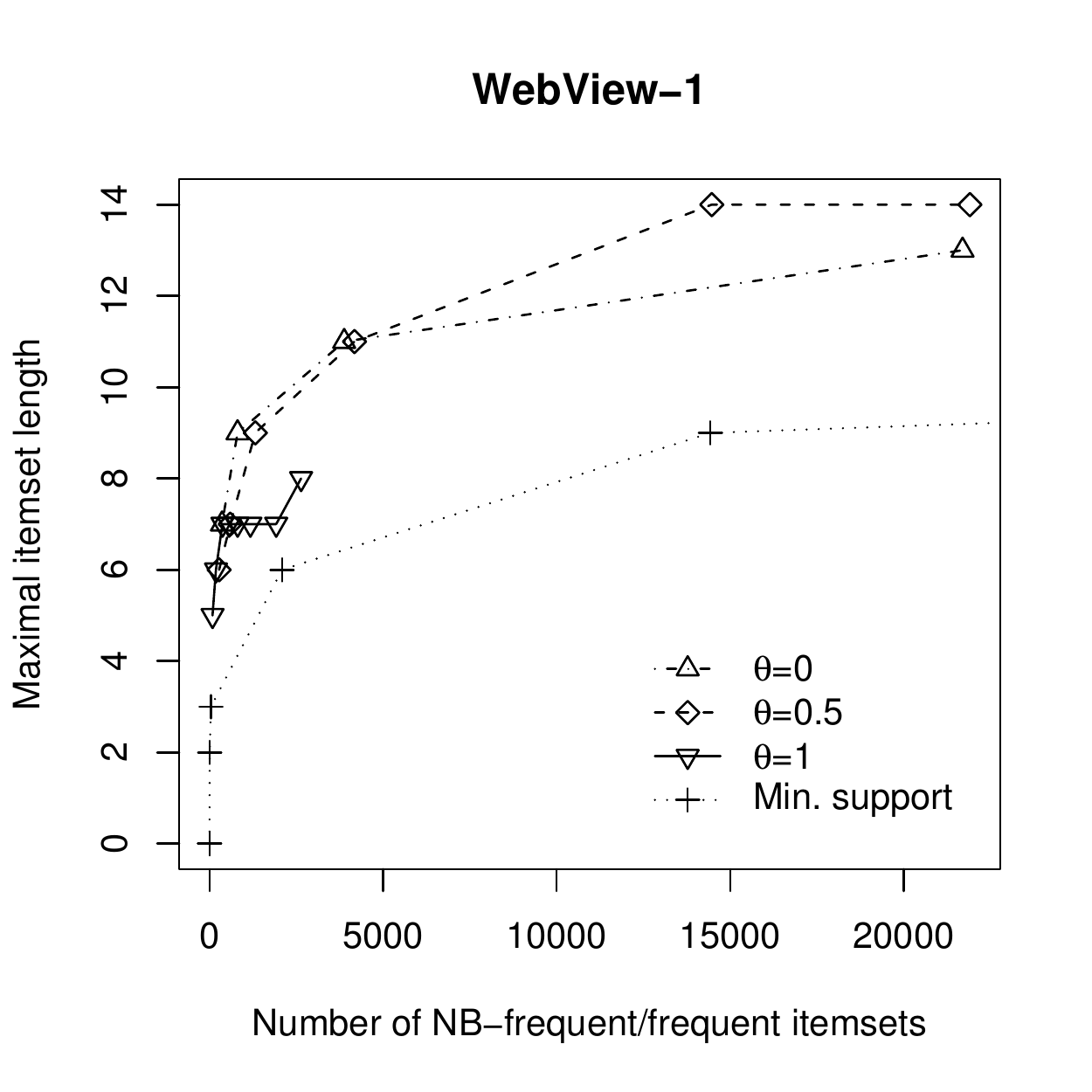}
\end{minipage}

\begin{minipage}[b]{.48\linewidth}
\includegraphics[scale=0.48]{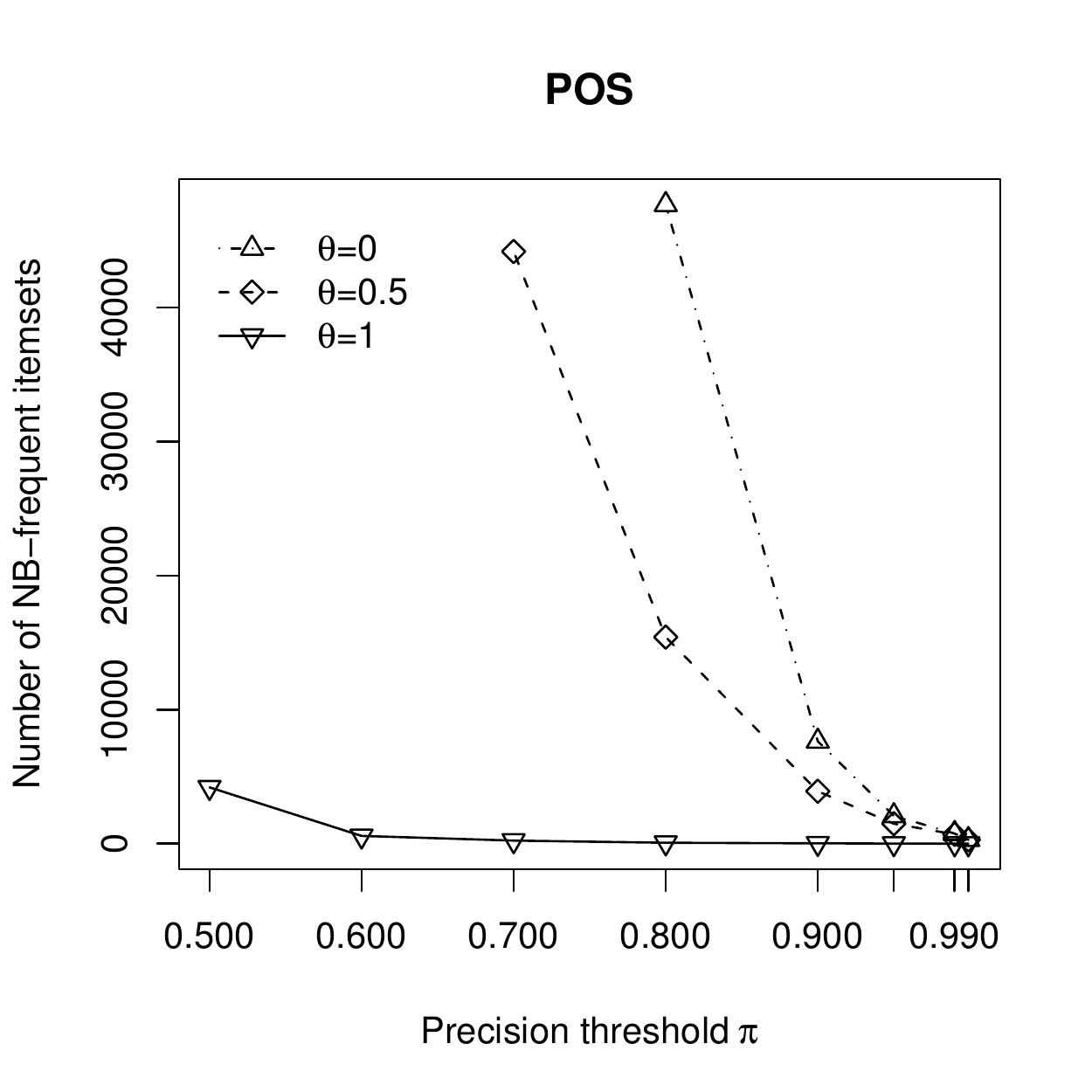}
\end{minipage}
\begin{minipage}[b]{.48\linewidth}
\includegraphics[scale=0.48]{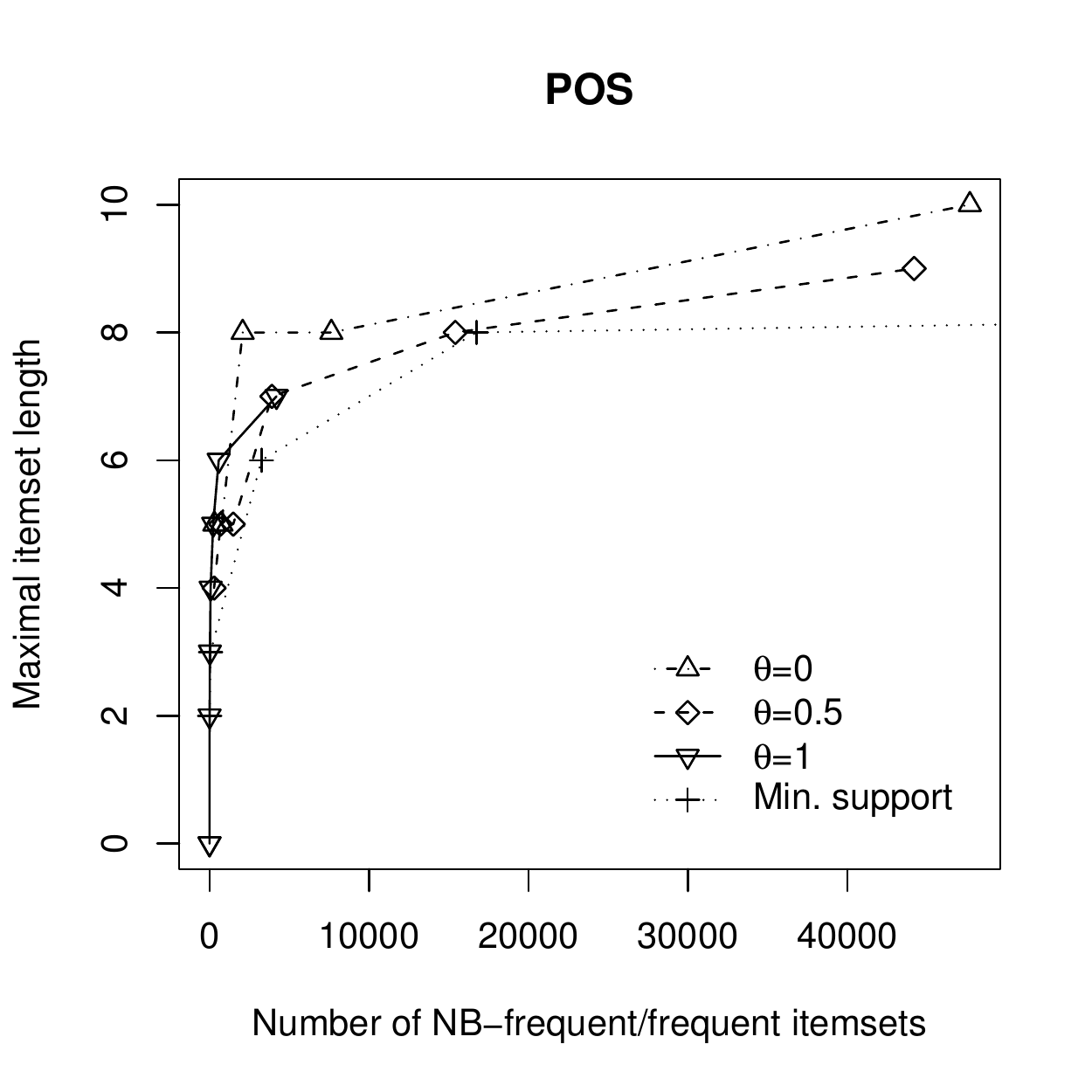}
\end{minipage}

\begin{minipage}[b]{.48\linewidth}
\includegraphics[scale=0.48]{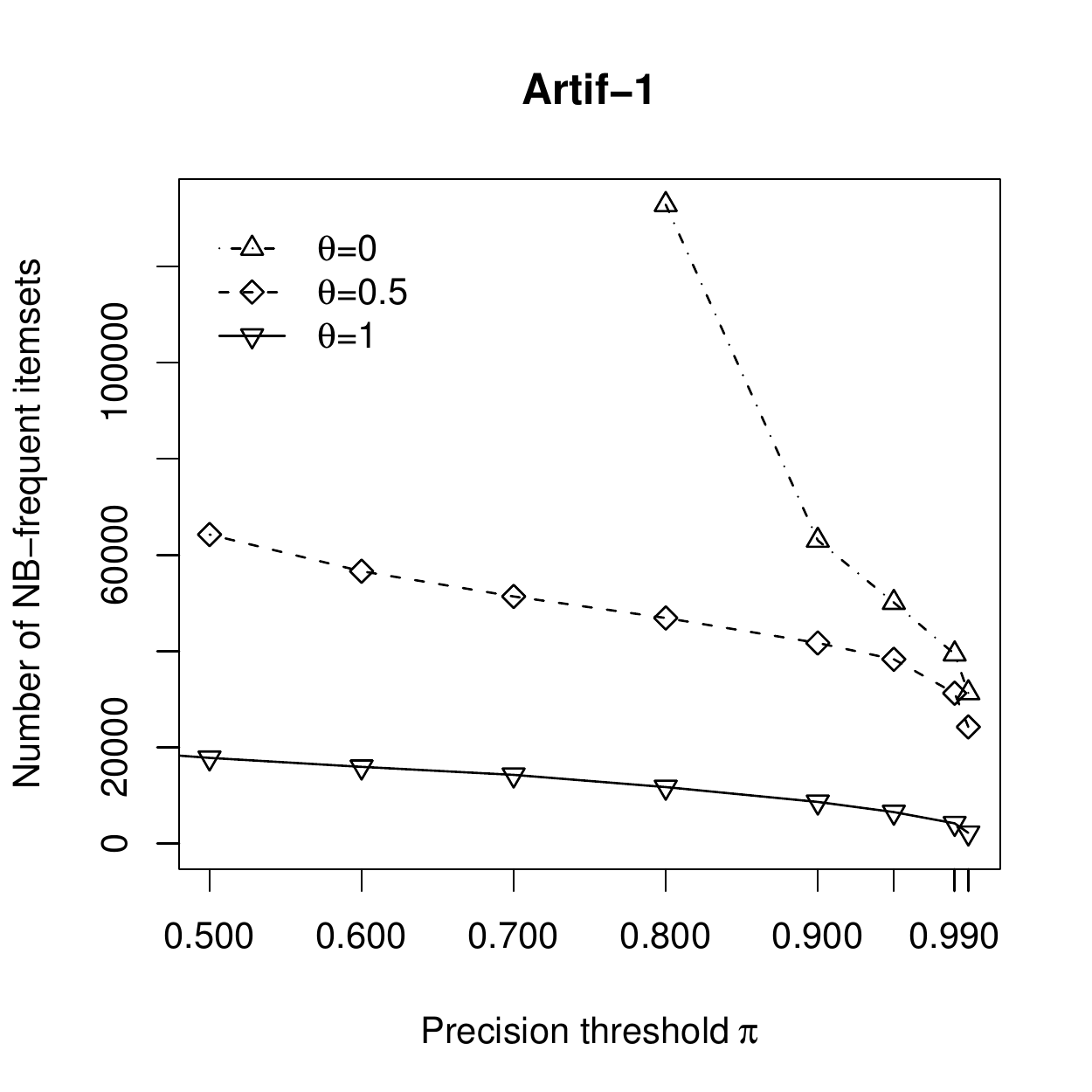}
\end{minipage}
\begin{minipage}[b]{.48\linewidth}
\includegraphics[scale=0.48]{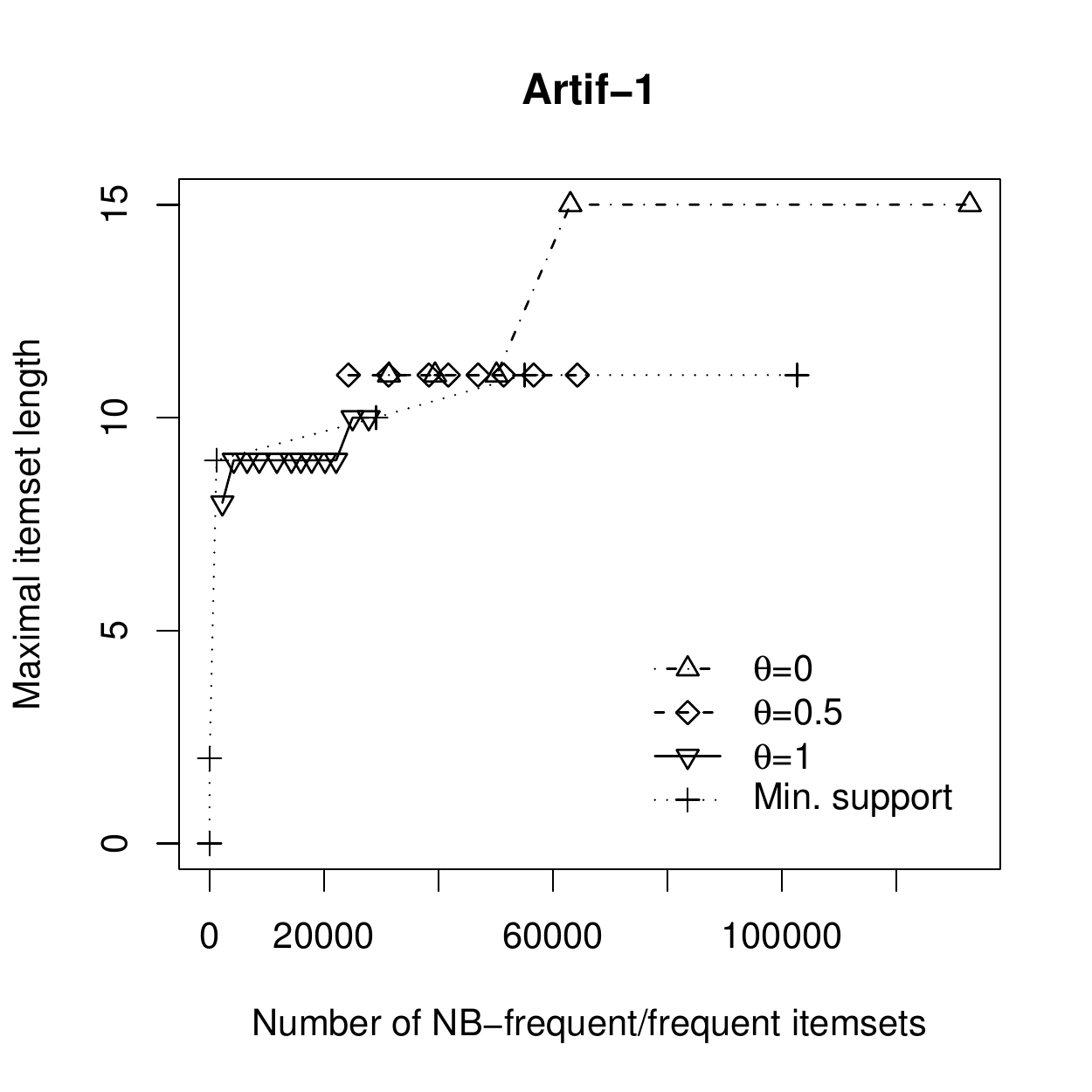}
\end{minipage}
\caption{Comparison of the number of generated NB-frequent itemsets 
for different parameter settings.\label{comp_patterngen}} 
\end{figure}

We report the influence of the different settings for $\theta$ and $\pi$ on the
three data sets already used in this paper in the plots in
Fig.~\ref{comp_patterngen}. In the left-hand side plots we see that by reducing
$\pi$ the number of generated NB-frequent itemsets increases for all settings
of $\theta$.  For the most restrictive setting $\theta=1$, pruning is maximal
and the number of NB-frequent itemsets only increases at a very moderate rate with
falling $\pi$.  For $\theta=0$, no pruning is conducted and the number of
NB-frequent itemsets explodes already at relatively high values of $\pi$.  At the
intermediate setting of $\theta=0.5$, the number of NB-frequent itemsets grows at
a rate somewhere in between the two extreme settings.  Although,
for the extreme settings all three data sets react similarly, for $\theta=0.5$
there is a clear difference visible between the real-world data sets and the
artificial data set.  While the magnitude of pruning for the real-world
data sets is closer to $\theta=0$, the magnitude for the artificial data set is
closer to $\theta=1$. Also, for the artificial data set we already find a
relatively high number of NB-frequent itemsets at $\pi$ near to one (clearly
visible for $\theta=0$ and $\theta=0.5$), a characteristic which the real-world
data sets do not show.  This characteristic is due to the way by which the used
generator produces the data set from maximal potentially large itemsets
(see~\cite{Agrawal1994}).

\begin{figure} 
\centering
\includegraphics[scale=0.6]{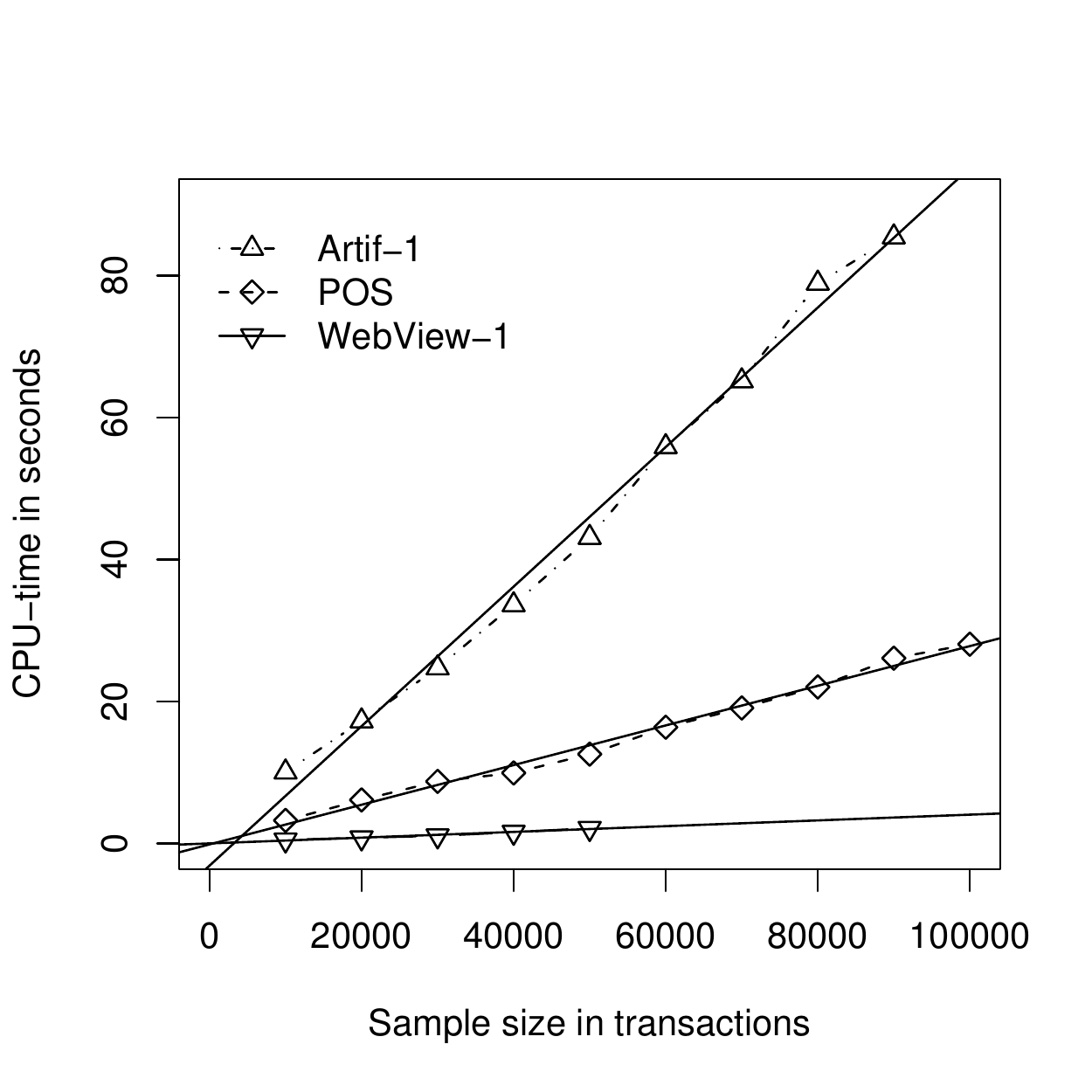}
\caption{Relationship between execution time and data set size for the setting
$\pi = 0.95$ and $\theta = 0.5$.\label{timing2}} 
\end{figure}

As for most other mining algorithms, the number of
generated itemsets has a direct influence on the 
execution time needed by the algorithm.
To analyze the influence of the growth of the number of 
NB-frequent itemsets with falling values for
parameter $\pi$, we
recorded the CPU time\footnote{We used a machine with two 
Intel Xeon processors (2.4 GHz) running Linux (Debian Sarge).  The algorithm
was implemented in JAVA and compiled using the gnu ahead-of-time
compiler gcj version 3.3.5.  CPU time was recorded using the time
command and we report the sum of user and system time.} 
needed by the algorithm for the data sets in Fig.~\ref{comp_patterngen}.
The results for the setting $\theta=0.5$ and the three data sets is
presented in Table~\ref{timing}. 
As for other algorithms, execution time mainly
depends on the search space size (given by the number of items)
and the structure (or sparseness) of the data set.
Compared to the other two data sets, WebView-1 has fewer items and is
extremely sparse with very short transactions (on average only
$2.5$ items).  Therefore, the algorithm needs to search through less
itemsets and takes less time (between 0.55 and 6.98 seconds for 
values of $\pi$ between 0.999 and 0.7).  Within each data set the 
execution time for
different settings of the parameter $\pi$ depends on how much of
the search space needs to be traversed.  
Since the traversed search space and the number of
generated NB-frequent itemsets is inversely related, 
the needed time grows close to linear with the number of
found NB-frequent itemsets (compare the execution times with the 
left-hand side plots in Fig.~\ref{comp_patterngen}).  
As for other algorithms, we can see from the pseudocode of the 
algorithm, that execution time is roughly linear 
in the number of transactions. This is supported by the experimental results
for different size samples from the three data sets displayed 
in Fig.~\ref{timing2}.

\begin{table} 
\caption{CPU-time in seconds to mine $20,000$
transactions of the data sets using $\theta=0.5$.\label{timing}}
\vspace{3mm}
\centering 
\begin{tabular}{cccc} 
\hline  
$\pi$ & {\bf WebView-1} & {\bf POS} & {\bf Artif-1} \\ 
\hline  
0.999 &0.55& 4.05&13.21 \\ 
0.99	&0.67& 4.85&15.03 \\
0.95	&0.92& 6.14&17.32 \\ 
0.9 	&1.61&12.38&18.27 \\
0.8	&3.88&36.90&19.28 \\ 
0.7   &6.98&80.66&20.81 \\ 
\hline 
\end{tabular} 
\end{table}

Next, we analyze the size of the accepted itemsets.  For comparison
we generated frequent itemsets using the implementations of
Apriori and Eclat by Christian Borgelt\footnote{Available at
http://fuzzy.cs.uni-magdeburg.de/\~{}borgelt/software.html}.  We
varied the minimum support threshold $\sigma$ between 0.1 and
0.0005.
These settings were found after some experimentation to work best for the
data sets.
In the plots to the right in Fig.~\ref{comp_patterngen} we show the
maximal itemset length by the number of accepted (NB-frequent or
frequent) itemsets for the data sets and the settings used in the
plot to the left.  Naturally, the maximal length grows for all
settings with the number of accepted itemsets which in turn grows
with a decreasing precision threshold $\pi$ or minimum support
$\sigma$.  For the real-world data sets, NB-DFS
tends to accept longer itemsets for the same number of accepted
itemsets than minimum support.  For the artificial data a clear
difference is only visible for the setting $\theta = 0$.

\begin{figure} 
\centering
\includegraphics[scale=0.6]{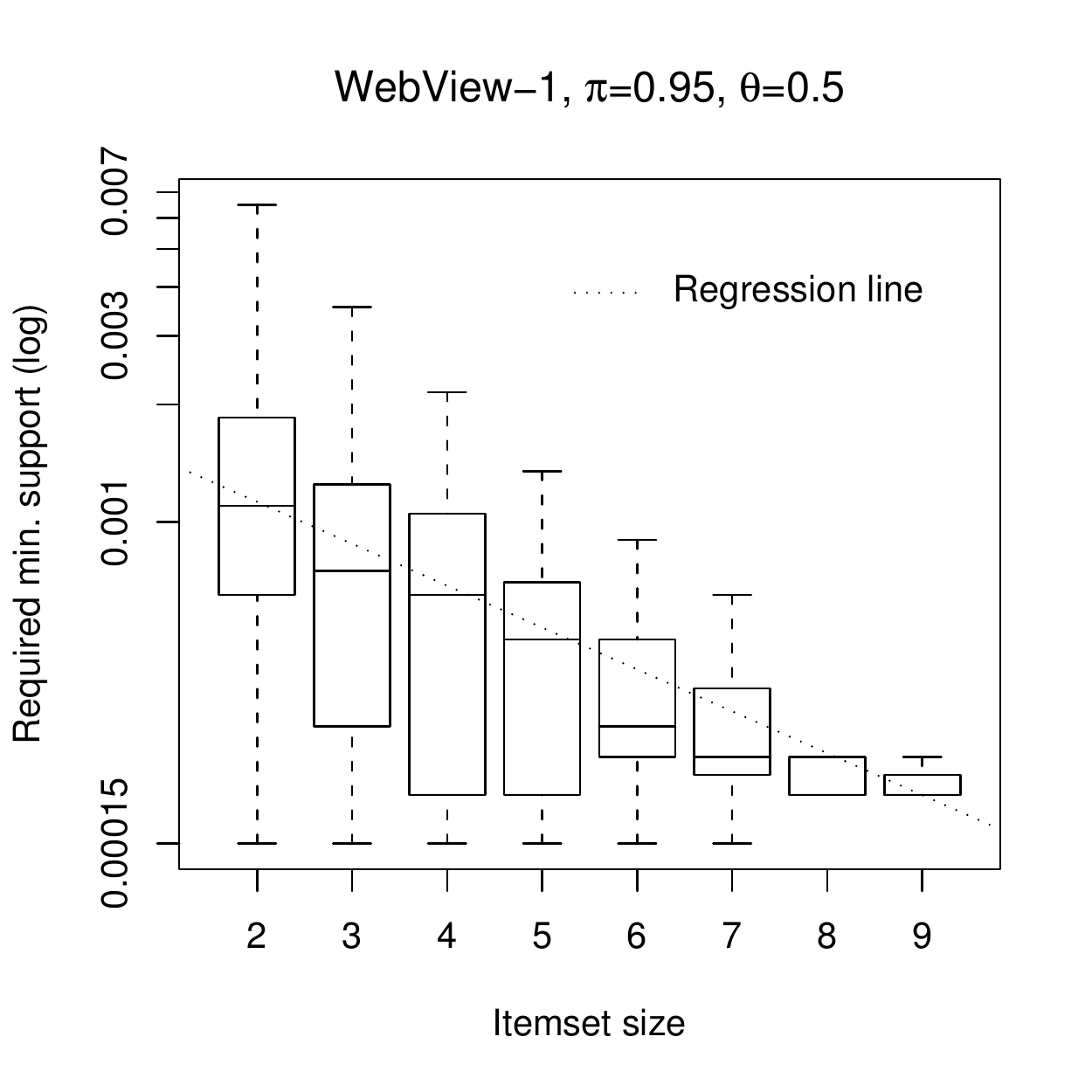}
\caption{Boxplot of the minimum
support required by NB-Select for the $1$-extensions of 
NB-frequent itemsets of different size.
\label{supportBYk}} 
\end{figure}

The longer maximal itemset size for the model-based algorithm is
caused by NB-Select's way of choosing an individual frequency
constraint for all $1$-extensions of an NB-frequent itemset.  
To analyze this
behavior, we look at the minimum supports required by NB-Select for
the data set WebView-1 at $\pi=0.95$ and $\theta=0.5$.  In
Fig.~\ref{supportBYk} we use a box-and-whisker plot to represent
the distributions of the minimum support thresholds required by
NB-Select for different itemset sizes.  The lines inside
the boxes represent the median required minimum supports, the box
spans from the lower to the upper quartile of the values, and the
whiskers extend from the minimum to the maximum.  The plot shows
that the required support falls with itemset size.

\cite{Seno2001} already proposed to reduce the
required support threshold with itemset size to improve the chances
of finding longer maximal frequent itemsets without being buried in
millions of shorter frequent itemsets. 
Instead of a fixed minimum support  they suggested 
using a minimum support function
which decreases with itemset size.  \cite{Seno2001} used in their
example a linear function together with an
absolute minimum, however, the optimal choice of a support function
and its parameters is still an open research question.  In contrast to
their approach, there is no need to specify such a function for the
model-based frequency constraint
since NB-Select automatically adjusts
support for all $1$-extensions of a NB-frequent itemset.
In Fig.~\ref{supportBYk} we see that the average 
required minimum support falls roughly at a constant rate 
with itemset size (the dotted straight line in the
plot represents the result of a linear regression on the logarithm of 
the required minimum supports). 
Reducing support with itemset size by a constant rate seems to be more 
intuitive than using a linear function.


\subsection{Effectiveness of Pattern Discovery} 
After we studied
the itemset generation behavior of the model-based algorithm 
and its ability to
accept longer itemsets than minimum support, 
we need to
evaluate if these additionally discovered itemsets represent
non-spurious associations in the database.  For the evaluation we need
to know what true associations exist in the data and then compare
how effective the algorithm is in discovering these itemsets.
Since for most real-world data sets the underlying associations are
unknown, we resort to artificial data sets, where the generation
process is known and can be completely controlled. 

To generate artificial data sets we use the popular generator developed by
\cite{Agrawal1994}.  To evaluate the
effectiveness of association discovery, we need to know all associations 
which were used to generate the data set.  In the original version
of the generator only the associations with the highest occurrence
probability are reported.  Therefore, we adapted the code of the
generator so that all used associations (called maximal potentially large
itemsets) are reported.  We generated
two artificial data sets using this modified generator.  
Both data sets consist of 
$|{\cal D}|=100,000$ transactions, the average 
transaction size is $|T|=10$,
the number of items is $N=1,000$, and for the correlation and
corruption levels we use the default values  ($0.5$ for both).

The first data set, Artif-1, represents the standard data set
T10I4D100K presented by~\cite{Agrawal1994} and which is used for
evaluation in many papers.
For this data set $|L|=2,000$ maximal potentially large itemsets
with an average size of $|I|=4$ are used.

For the second data set, Artif-2, we decrease the average association 
size to $|I|=2$. This will produce more maximal potentially large
itemsets of size one.  These $1$-itemsets are not
useful associations since they do not provide information about dependencies
between items.
They can be considered noise in the generated database and, therefore,  
make finding longer associations more difficult.  
A side effect of reducing the
average association size is that the chance of using longer
maximal potentially large itemsets for the database generation
is reduced.
To work against this effect, we double their
number to $|L|=4,000$.  

For the experiments,
we use for both data sets the first $20,000$ transactions for
mining associations. To analyze how the effectiveness is
influenced by the data set size, we also report results for sizes $5,000$
and $80,000$ for Artif-2. For the model-based algorithm we
estimated the parameters of the model from the data sets and then 
mined NB-frequent itemsets
with the settings $0, 0.5$ and $1$ for $\theta$.
For each of the three settings for $\theta$, 
we varied the parameter $\pi$ between 0.999 and 0.1 (0.999,
0.99, 0.95, 0.9, 0.8 and in 0.1 steps down to 0.1).
Because of combinatorial explosion discussed in the previous section, 
we only used $\pi \ge 0.5$ for $\theta=0.5$ and $\pi \ge 0.8$ for $\theta=0$.

For comparison with existing methods we 
mined 
frequent itemsets at minimum support levels
between 0.1 and 0.0005 (0.01, 0.005, 0.004, 0.003, 0.002, 0.0015,
0.0013, 0.001, 0.0007, and 0.0005).  
And as a second benchmark we
generated itemsets using all-confidence. 
We
varied the threshold on all-confidence between 0.01 and 0.6 (0.6,
0.5, 0.4, 0.3, 0.2, 0.1, 0.05, 0.04, 0.03, 0.02, 0.01).  The used
minimum support levels and all-confidence thresholds were found
after some experimentation to cover a wide area of the possible
true positives/false positives combinations for the data sets.

To compare the ability to discover associations which were used 
to generate the artificial data sets, we
counted the true positives (itemsets and their subsets 
discovered by the
algorithm which were used in the data set generation process) and
false positives (mined itemsets which were not used in
the data set generation process).  
This information together with
the total number of all positives in the database (all itemsets used to
generate a data set) is used to 
calculate precision (the ratio
of the number of true positives by the number of all instances
classified as positives)
and recall (the ratio of the
number of true positives by the total number of positives in the data).
Precision/recall
plots, a common evaluation tool in information retrieval
and machine learning,
are then used to
visually inspect the algorithms'
effectiveness over their parameter spaces.

\begin{figure} 
\begin{minipage}[t]{.48\linewidth}
\includegraphics[scale=0.48]{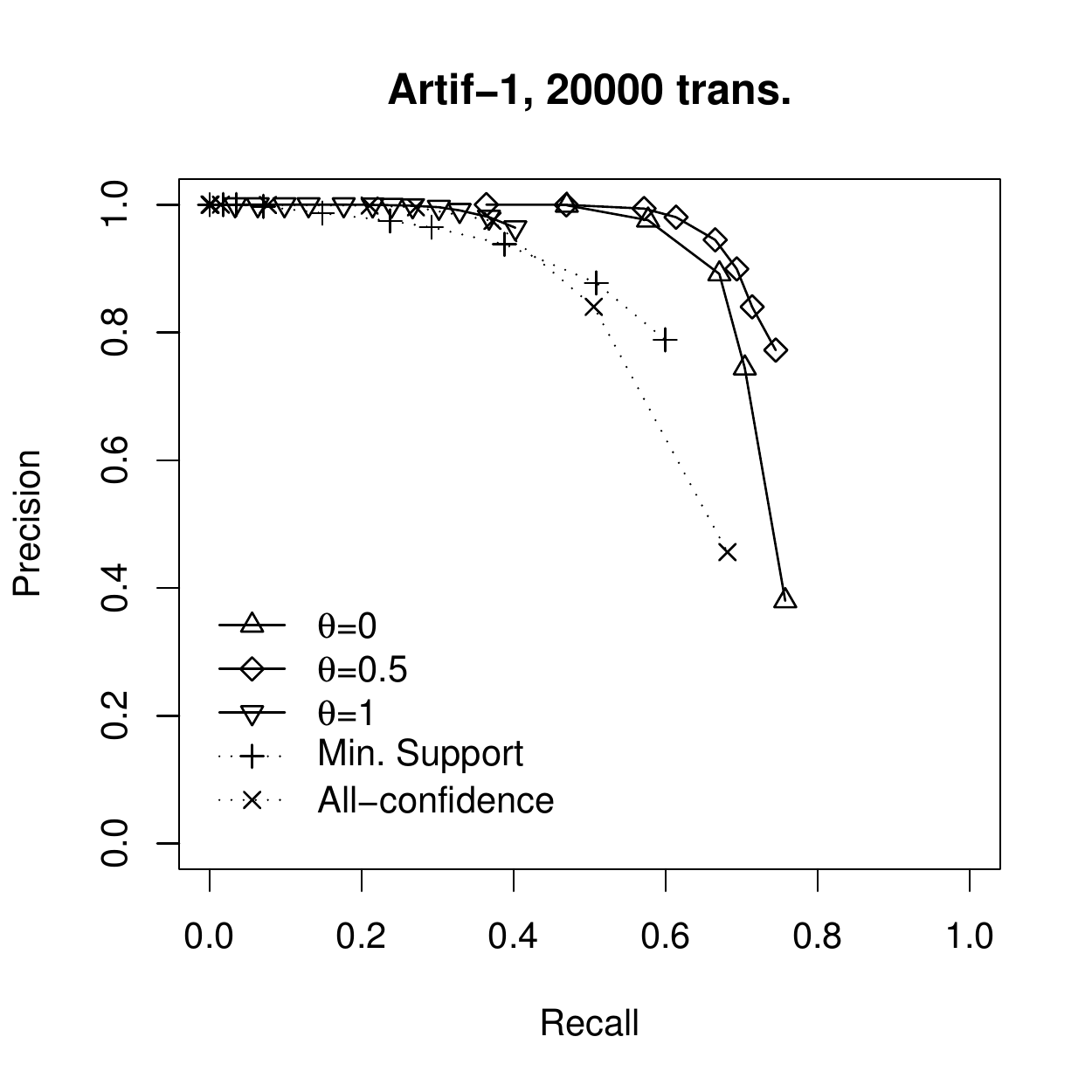}
\end{minipage}
\begin{minipage}[t]{.48\linewidth}
\includegraphics[scale=0.48]{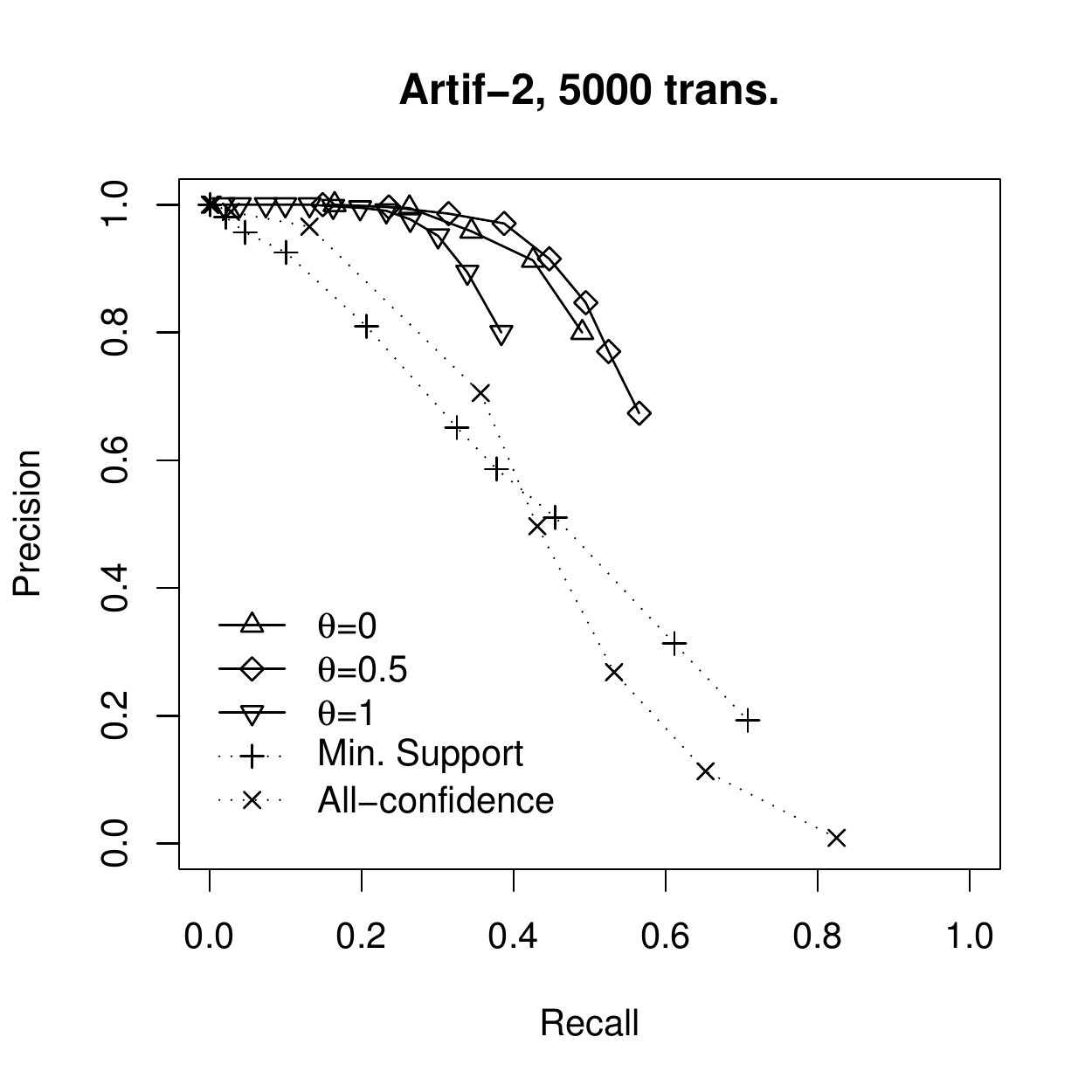}
\end{minipage}

\begin{minipage}[t]{.48\linewidth}
\includegraphics[scale=0.48]{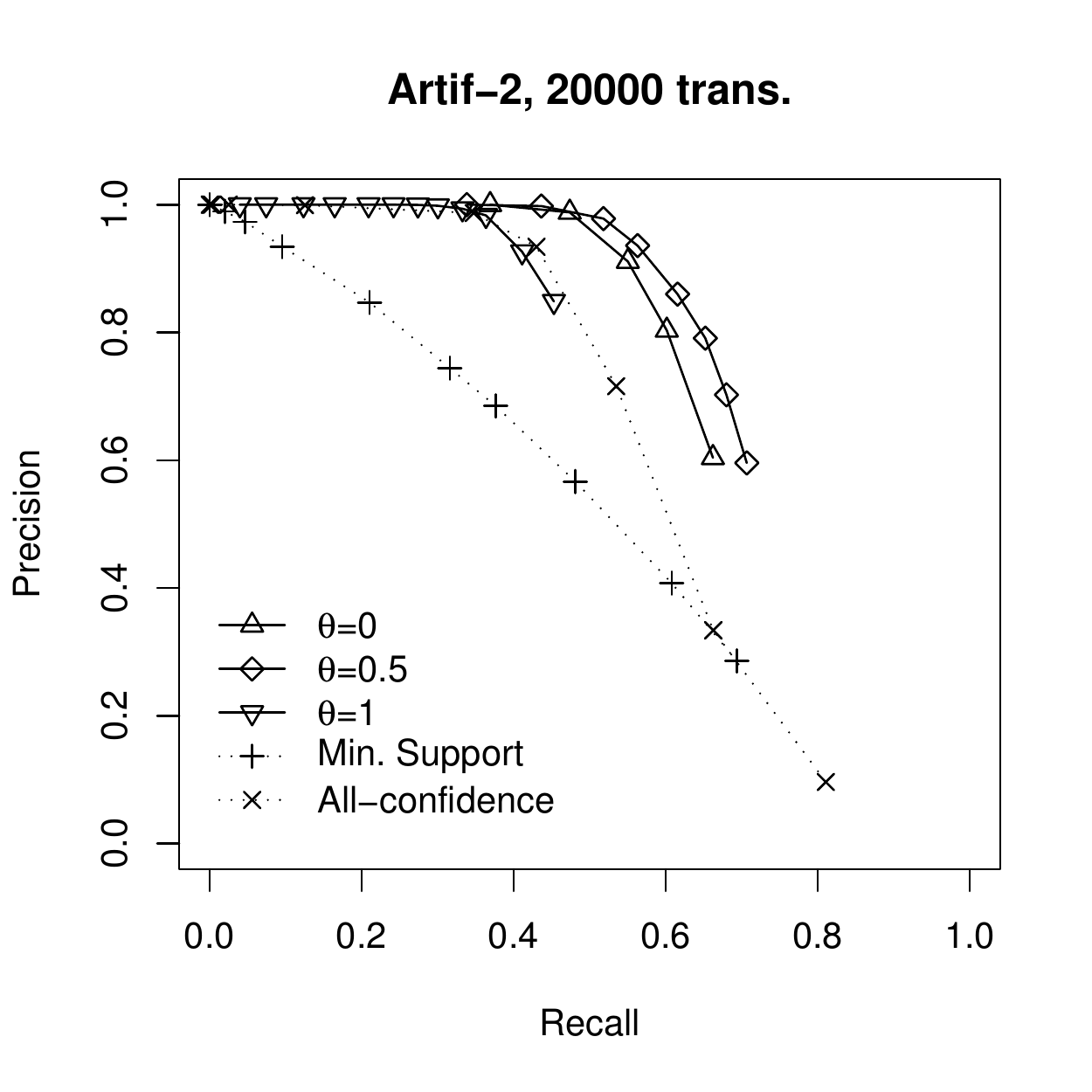}
\end{minipage}
\begin{minipage}[t]{.48\linewidth}
\includegraphics[scale=0.48]{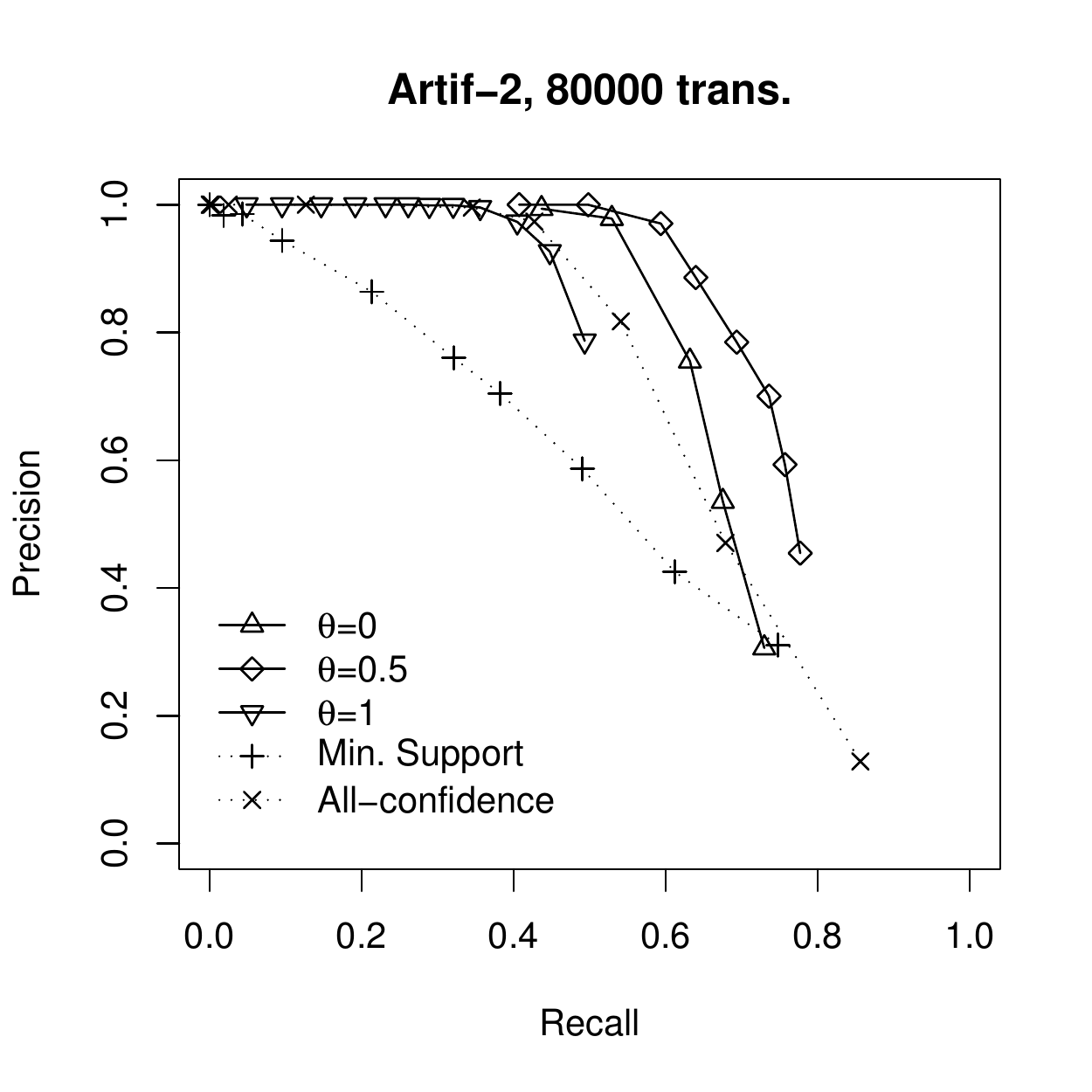}
\end{minipage}
\caption{Comparison of the effectiveness of the model-based 
constraint with different settings for $\theta$. \label{comp_fig}} 
\end{figure}

In Fig.~\ref{comp_fig}, we inspect the effectiveness 
of the algorithm
using the three settings for $\theta$.
For comparison we add the
precision/recall curves for minimum support and all-confidence.
The top right
corner of each precision/recall plot represents the optimal
combination where all associations are discovered ($\mathrm{recall}=1$)
and no false itemset is selected ($\mathrm{precision}=1$).  Curves
that are closer to the top right corner represent better retrieval
effectiveness.  

The precision/recall plots show that with
$\theta=1$ and $\pi \ge 0.5$ reachable recall is comparably low,
typically smaller than $0.5$, while precision is always high.  On the data
sets with 20,000 transactions it shows similar effectiveness as
all-confidence.  However, it outperforms all-confidence
considerably on the small data set (Artif-2 with 5,000
transactions) while it is outperformed by all-confidence on the
larger data set (Artif-2 with 80,000 transactions).  This
observation suggests that, if only little data is available, 
the additional knowledge of the structure
of the data is more helpful.

With $\theta=0$, where the generation is least strict, 
the algorithm reaches higher recall but
precision deteriorates considerably with increased recall.  The
effectiveness is generally better than minimum support and
all-confidence.  Only for settings with very low values for $\pi$,
precision degrades so strongly that its effectiveness is worse than
minimum support and all-confidence.  
This effect can be seen in Fig.~\ref{comp_fig} 
for  data set
Artif-2 with 80,000 transactions.

The model-based algorithm with $\theta=0.5$ performs overall the
best with high recall while loosing less precision than $\theta=0$.  Its
effectiveness clearly beats minimum support, all-confidence, and the
model based algorithm with settings $\theta=0$ and $\theta=1$ on
all data sets.

Comparing the two precision/recall plots for the data sets 
with 20,000 transactions in Fig.~\ref{comp_fig}
shows that the results of the model-based
constraint (especially for $\theta=0.5$) dependent less on the
structure and noise of the data set.  To quantify this finding, we
calculate the relative differences between the resulting precision
and recall values for each parameter setting of each algorithm.  In
Table~\ref{comp_precrec} we present the average of the relative
differences per algorithm.  While precision differs for support
between the two data sets on average by about 30\%, all-confidence
exhibits an average difference of nearly 80\%.  
Both values clearly indicate, that the optimal choice of
the algorithms' parameters differs significantly 
for the two data sets.
The model-based
algorithm only differs by less than 20\%, and with $\theta=0.5$ the
precision difference is only about 7\%.  This suggests that setting
an average value for $\pi$  (e.g., 0.9) will produce reasonable
results independently of the data set.  The user only needs to
resort to experimentation with different settings for the parameter,
if she needs to optimize the results.

\begin{table} 
\caption{Relative differences of precision and recall
between the results from data sets Artif-1 and Artif-2 (both with
20,000 transactions).\label{comp_precrec}} 
\vspace{3mm}
\centering
\begin{tabular}{ccc} 
\hline  
	& Precision & Recall \\ 
\hline  
$\theta=1$&	15.94\%	& 1.54\% \\
$\theta=0.5$& 6.86\% &	8.32\% \\ 
$\theta=0$& 16.88\% &	14.09\% \\ 
Min. support&	30.65\%	& 23.37\% \\ 
All-confidence& 79.49\%	& 17.31\% \\
\hline  
\end{tabular} 
\end{table}

For an increasing data set size (see Artif-2 with
80,000 transactions in Fig.~\ref{comp_fig}) and for the model-based 
algorithm at a set
$\pi$, recall increases while at the same time precision decreases.
This happens because with more available data NB-Select's
predictions for precision get closer to the real values.
In Table~\ref{comp_actualprec}, we summarize the actual precision of
the mined associations with $\theta=0.5$ at different settings for
the precision threshold.
The close agreement between the columns indicates that, with enough
available data, the set threshold on the predicted precision
gets close to the actual
precision of the set of mined associations.  This is an important
property of the model-based constraint, since it makes the precision
parameter easier to understand and set for the person who applies
data mining.  While suitable thresholds on measures as support and
all-confidence are normally found for each data set by
experimentation, the precision threshold can be set with a 
needed minimal precision (or maximal
acceptable error rate) for an application in mind.

\begin{table}
\caption{Comparison of the set precision threshold $\pi$ and the
actual precision of the mined associations for $\theta=0.5$ on data
set Artif-2 with 80,000 transactions.\label{comp_actualprec}}
\vspace{3mm}
\centering 
\begin{tabular}{cc} 
\hline  
$\pi$ & precision \\
\hline  
0.999 & 1.0000000 \\
0.990 & 0.9997855 \\
0.950 & 0.9704649 \\
0.900 & 0.8859766 \\
0.800 & 0.7848500 \\ 
0.700 & 0.7003764 \\ 
0.600 & 0.5931635 \\
0.500 & 0.4546763 \\
\hline  
\end{tabular}
\end{table}

A weakness of precision/recall plots and many other ways to
measure accuracy is that they are only valid for comparison under
the assumption of uniform misclassification cost, i.e., the error
cost for false positives and false negatives are equal.
A representation that does
not depend on uniform misclassification cost
is the {\em Receiver Operator Characteristics graphs} (ROC graphs)
used in machine learning to compare classifier
accuracy~\citep{Provost1997}. 
It is independent of class distribution
(proportion of true positives to true negatives) and the
distribution of misclassification costs.  
A ROC graph is a plot
with the {\em false positive rate} on 
the x-axis and the {\em true positive rate} 
on the y-axis and represents the possible error
trade-offs for each classifier.
If a classifier can be parametrized, the points obtained using
different parameters can be connected by a line 
called a {\em ROC curve.} 
If all points of one classifier are superior to the points
of another classifier, the first classifier is said to dominates the latter
one. This means that for all possible cost and class distributions,
the first classifier can produce better results.  We also examined
ROC curves (omitted here due to space restrictions) 
for the data sets producing basically the same results
as the precision/recall plots. The model-based frequency
constraint with $\theta = 0.5$ clearly dominates all other settings as well
as minimum support and all-confidence.
%

The results from artificial data sets presented here might not
carry over 100\% to real-world data sets.  However, the
difference between the effectiveness of the model-based constraint 
with $\theta=0.5$ and 
minimum support or all-confidence is so big, that also on
real-world data a significant improvement can be expected.

\section{Conclusion\label{conclusion}}
The contribution of this paper is that we presented a model-based
alternative to using a single, user-specified minimum support
threshold for mining  associations in transaction data.  
We extended a simple and robust stochastic mixture model (the NB
model) to develop a baseline model for incidence counts
(co-occurrences of items) in the database. The model is easy 
to fit to data and explains co-occurrences counts between independent items.
Together with a user-specified precision threshold,
a local frequency constraint (support threshold)
for all $1$-extensions of an itemset can be found.
The precision threshold represents the predicted error rate in the 
mined set of associations and, therefore, 
it is easy to specify by the user with the requirements of a
specific application in mind.

Based on the developed model-based frequency constraint, we introduced the
notion of NB-frequent itemsets
and presented a prototypical mining algorithm
to find all NB-frequent itemsets in a database.  
Although the definition of NB-frequency, which
is based on local frequency constraints,  does not provide the
important downward closure property of support, we showed how the
search space can be adequately reduced to make efficient mining 
possible.

Experiments showed that the model-based frequency constraint
automatically reduces the average needed frequency (support) with
growing itemset size.  Compared with minimum support it tends to be more
selective for shorter itemsets while still accepting longer itemsets with
lower support.  This property reduces the problem of being buried
in a great number of short itemsets when using a relatively low
threshold in order to also find longer itemsets.

Further experiments on artificial data sets indicate that the
model-based constraint is more effective in finding
non-spurious associations.
The largest improvements were found for noisy data sets or when only
a relatively small database is available.
These experiments also show that the precision parameter of the
model-based algorithm depends less than support or any-confidence
on the data set. This is a huge advantage and reduces the need for
time-consuming experimentation with different parameter settings
for each new data set.

Finally, it has to be noted that the model-based constraint
developed in this paper can only be used for databases which are
generated by a process similar to the developed baseline model.
The developed baseline is a robust and reasonable model for most
transaction data (e.g., point-of-sale data).  For other types of
data, different baseline models can be developed and can then
be incorporated in mining algorithms following the outline of this
paper.


\section*{Acknowledgments}

The author wishes to thank Blue Martini Software
for contributing the KDD Cup 2000 data, Ramakrishnan Srikant from
the IBM Almaden Research Center for making the code for the
synthetic transaction data generator available, and to Christian
Borgelt for the free implementations of Apriori and Eclat.  

The author especially
thanks Andreas Geyer-Schulz and Kurt Hornik for the
long discussions on modeling transaction data and the anonymous referees
for their valuable comments.

%
\bibliographystyle{apalike}
\bibliography{recommender,info_goods,buying_behavior,association_rules,stat,hahsler,association_applications}
\end{document}